\documentclass[usenatbib]{emulateapj}
\usepackage{graphicx}
\usepackage[flushleft]{threeparttable}
\usepackage[usenames,dvipsnames,svgnames,table]{xcolor}
\usepackage{hyperref}
\definecolor{darkblue}{rgb}{0.0,0.0,0.3}
\hypersetup{colorlinks,breaklinks,
            linkcolor=darkblue,urlcolor=darkblue,
            anchorcolor=darkblue,citecolor=darkblue}
\usepackage{amsmath,amssymb}
\usepackage{amsmath}

\usepackage{natbib}
\begin{document}

\def\etal{et al.\ \rm}
\def\ba{\begin{eqnarray}}
\def\ea{\end{eqnarray}}
\def\etal{et al.\ \rm}
\def\Fdw{F_{\rm dw}}
\def\Tex{T_{\rm ex}}
\def\Fdis{F_{\rm dw,dis}}
\def\Fnu{F_\nu}

\newcommand\cmtrr[1]{{\color{red}[RR: #1]}}
\newcommand\cmtla[1]{{\color{blue}[LA: #1]}}


\title{Disk Accretion Driven by Spiral Shocks}

\author{Lev Arzamasskiy\altaffilmark{1} \& Roman R. Rafikov\altaffilmark{2,3}}
\altaffiltext{1}{Department of Astrophysical Sciences, 
Princeton University, Ivy Lane, Princeton, NJ 08540; leva@astro.princeton.edu}
\altaffiltext{2}{Centre for Mathematical Sciences, Department of Applied Mathematics and Theoretical Physics, University of Cambridge, Wilberforce Road, Cambridge CB3 0WA, UK}
\altaffiltext{3}{Institute for Advanced Study, Einstein Drive, Princeton, NJ 08540}


\begin{abstract}
Spiral density waves are known to exist in many astrophysical disks, potentially affecting disk structure and evolution. We conduct a numerical study of the effects produced by a density wave, evolving into a shock, on the characteristics of the underlying disk. We measure the deposition of angular momentum in the disk by spiral shocks of different strength and verify the analytical prediction of Rafikov (2016) for the behavior of this quantity, using shock amplitude (which is potentially observable) as the input variable. Good agreement between the theory and numerics is found as we vary shock amplitude (including highly nonlinear shocks), disk aspect ratio, equation of state, radial profiles of the background density and temperature, and pattern speed of the wave. We show that high numerical resolution is required to properly capture shock-driven transport, especially at low wave amplitudes. We also demonstrate that relating local mass accretion rate to shock dissipation in rapidly evolving disks requires accounting for the time-dependent contribution to the angular momentum budget, caused by the time dependence of the radial pressure support. We provide a simple analytical prescription for the behavior of this contribution and demonstrate its excellent agreement with the simulation results. Using these findings we formulate a theoretical framework for studying one-dimensional (in radius) evolution of the shock-mediated accretion disks, which can be applied to a variety of astrophysical systems.
\end{abstract}




\section{Introduction.}  
\label{sect:intro}


Evolution of astrophysical disks is caused by the redistribution, gain or loss of angular momentum of the disk fluid, ultimately resulting in accretion onto the central object. Angular momentum of the fluid elements can change as a result of internal or external stresses. The former usually involve turbulence produced by operation of some instability in the disk. The most common example of such an instability is the magneto-rotational instability (MRI, \citealt{Velikhov,Chandrasekhar,Balbus}), but it may fail to operate in weakly ionized protoplanetary disks \citep{Turner}. Other processes, such as the gravitoturbulence \citep{Gammie,Rafikov15}, vertical shear instability \citep{Urpin,Stoll}, etc. have been proposed to explain the observed evolution of cold protoplanetary disks. External stresses can be naturally exerted on the disk via outflows \citep{BP,Konigl}, resulting in the loss of angular momentum from the system and mass accretion. 

Another potential driver of the disk evolution could be the global density waves, ubiquitous in astrophysical disks. Previously, observations using the technique of Doppler tomography revealed presence of spiral waves excited by the gravity of the donor star in accretion disks of cataclysmic variables \citep{Spiral,Marsh}. Global asymmetric emission pattern has also been found in a compact gaseous debris disk of a metal-rich white dwarf SDSS J122859.93+104032.9 \citep{Manser}. Recent observations of protoplanetary disks using direct imaging in optical and near-IR \citep{Benisty}, as well as sub-mm interferometry with {\it ALMA} \citep{ALMA_spirals,Perez}, show spiral arms to be a common phenomenon. They are believed to be excited by the gravity of embedded massive planets  \citep{Boccaletti,Collins,Garufi,Grady,Muto}, stellar companions \citep{Wagner,Dong_star,Benisty16}, or by gravitational instability in a massive disk \citep{Meru}. Global spirals can also be driven by the non-uniform illumination of the disk by the central star \citep{Montesinos}, caused by the shadowing of the outer disk by an inclined inner disk \citep{Casassus,Marino}.

Spiral density waves induced by the gravity of a massive companion have also been routinely seen in numerical studies of disks \citep{Sawada1,Sawada2}, both in pure hydro \citep{Savonije} and in the MHD simulations \citep{Ju}. Numerical simulations reveal that the high-amplitude spiral density waves rapidly evolve into {\it spiral shocks}, as the non-linear effects play important role in their propagation \citep{Rafikov02,Dong_star,Dong_spiral,Zhu}. Recent work of \citet{Ju,Ju2017} on cataclysmic variables, subsequently extended to circumplanetary disks \citep{ZhuCP}, showed that spiral shocks can be important contributors to the angular momentum and mass transport even in well ionized disks with fully developed MRI. A similar conclusion was reached by \citet{Jiang} in their study of super-Eddington accretion around supermassive black holes.

On the theoretical side, properties of spiral shocks in accretion disks were first considered under the assumption of self-similarity, starting with the work of \citet{Spruit87}, \citet{SpruitSawada}, and \citet{Larson}. More recently, self-similar shock solutions in disks were constructed by \citet{Hennebelle} with the goal of explaining spiral waves observed in simulations of disks with the envelope infall \citep{Lesur}. A limitation of the self-similar shock solutions is that they apply only to special disk models with certain scale-invariant profiles of the surface density. Real astrophysical disks need not be described by these idealized models. 

\citet{GR01} explored the evolution of weakly nonlinear density waves excited in the protoplanetary disks by the low-mass planets. Their work, cast in the local, shearing sheet approximation, provided a formalism for describing the wave propagation, shock formation and its subsequent dissipation. \citet{GR01} have shown, in particular, that an ensemble of density waves launched by a population of planets can lead to evolution of a protoplanetary disk on $\sim$Myr timescale. This work has been subsequently extended by  \citet{Rafikov02} to {\it global} density waves (not limited to the vicinity of the perturber), making wave evolution formalism applicable to disks with arbitrary distributions of temperature and density. This framework was then used in \citet{Rafikov02b} to consider gap opening by planets via the nonlinear dissipation of their density waves. Numerical work of \citet{Li2009}, \citet{Yu}, \citet{DongNL}, \citet{Duffell}, \citet{Li2013}, \citet{Fung} have largely confirmed these analytical results.

More recently, \citet{Rafikov16} analytically calculated the effect of a spiral shock of {\it arbitrary} strength (i.e. not only weakly nonlinear) on the disk, through which it propagates. He showed that shocks of even moderate strength can easily drive mass accretion through the disk at rates comparable to $\dot M$ due to internal stresses (in agreement with the simulations of \citealt{Ju}). These analytical predictions for the shock-driven $\dot M$ were confirmed by \citet{Ryan}, who characterized shock strength by measuring the entropy jump across its front, as suggested in \citet{Rafikov16}. 

Observationally, the most accessible characteristic of the shock is not the entropy jump, but the {\it density contrast} across the shock. Even that quantity may be difficult to measure, especially in scattered light observations of protoplanetary disk, when the brightness contrast at the shock is more sensitive to the corrugations of the disk surface than to the variations of the surface density. Nevertheless, numerical simulations with radiation transfer post-processing can be used to calibrate the relation between the brightness contrast and the density jump, such as recently done by \citet{DongFung}. Also, optically thin sub-mm observations using {\it ALMA} should be able to provide a direct probe of the density contrast across the shock. 

This observational connection provides a motivation for our present work. Here we seek to numerically verify the analytical predictions of \citet{Rafikov16} formulated in terms of the (observable) density contrast across the shock. This provides a new way of understanding the shock-driven disk evolution, which is complementary to that used by \citet{Ryan}. 

Our work is structured as follows. In \S \ref{sect:main} we summarize the analytical background for our calculations, while in \S \ref{sect:num_setup} we describe our numerical setup. The typical results of our simulations are discussed in \S \ref{sect:results}, and in \S \ref{sect:theory} we provide an exhaustive comparison between the numerical results and theoretical predictions of \citet{Rafikov16}. The emergence of the secondary shocks in our simulations and their effect on disk evolution are described in \S \ref{sect:nontrivial}. In \S \ref{sect:time-dep} we provide an analytical prescription for computing the time-dependent angular momentum contribution emerging in our simulations, which allows us to formulate a general framework for following the shock-driven disk evolution in \S \ref{sect:formalism}. We discuss and summarize our results in \S \ref{sect:disc} and \ref{sect:summary}, respectively.


\section{Basic equations.}  
\label{sect:main}


We consider a two-dimensional (2D) fluid disk and characterize its properties in polar coordinates $R,\phi$ via the surface density $\Sigma$ and radial and azimuthal velocities $v_R$ and $v_\phi$. We are interested in the effects of a spiral shock propagating through the disk on the angular momentum and mass transport. For simplicity, the disk is assumed to be unmagnetized and we neglect the gravitational effect of any external perturbers as well as the disk self-gravity. This does not mean that the density wave propagating through the disk cannot be launched via the gravitational coupling to an external perturber such as a planet or stellar companion. This assumption simply implies that we are studying the disk far from the wave excitation region, so that the torque induced on the density wave by the perturber's gravity can be ignored. 

Angular momentum conservation in a disk can be described quite generally via the  equation \citep{BalPap,Ju}
\ba
\partial_t\langle\Sigma R v_\phi\rangle+\frac{1}{R}\partial_R\langle R^2\Sigma v_R v_\phi\rangle=0, 
\label{eq:AMcons1}
\ea
where $\langle ...\rangle$ implies integration over $\phi$, 
$\langle f\rangle\equiv\int_0^{2\pi}f(R,\phi)d\phi$ for any function $f(R,\phi)$. As the fluid motion does not depart too far from axisymmetry, it is convenient to define azimuthal velocity perturbation $\delta v_\phi\equiv v_\phi-\Omega(R)R$, where $\Omega$ is the angular velocity in the absence of perturbation. Then one can easily show that the equation (\ref{eq:AMcons1}) reduces to
\ba
R^2\partial_t\langle\Sigma \delta v_\phi\rangle-\dot M\partial_R l = -\partial_R\left(R^2 T_{R\phi}\right), 
\label{eq:AMcons}
\ea
where $\dot M\equiv -\langle\Sigma R v_R\rangle$ is the mass accretion rate (defined to be positive for {\it inflow}), $l\equiv \Omega(R)R^2$ is the specific angular momentum, and $T_{R\phi}\equiv \langle\Sigma  v_R\delta v_\phi\rangle$ is the Reynolds stress. 

Note that the choice of $\Omega$ is to some extent arbitrary --- changing $\Omega$ affects both the first term in the left hand side and the right hand side ($T_{R\phi}$ depends on $\delta v_\phi$) of equation (\ref{eq:AMcons}) in such a way that the relation (\ref{eq:AMcons1}) always holds true. It is customary to assume $\Omega(R)=\Omega_{\rm K}(R)=\left(GM_\star/R^3\right)^{1/2}$ \citep{BalPap,Ju}, and we will adopt this convention in our numerical work as well.

Disks evolving predominantly under the action of internal stress feature non-zero $T_{R\phi}$, which drives their evolution and mass transport (i.e. makes $\dot M$ non-zero) as demonstrated by equation (\ref{eq:AMcons}). Internal stress can be provided by turbulence driven by some instability in the disk, e.g. MRI. 

Situation is different in disks evolving under the action of the global spiral shocks. In this case the angular momentum is {\it directly injected} into the disk at the shock fronts. Between the shock fronts the angular momentum is conserved in the absence of internal stresses. Even though such pattern of angular momentum injection is highly localized in azimuthal angle, the long-term effect of the shocks on the disk fluid can still be described via the azimuthally averaged quantities. In particular, we can assume that shock damping transfers the momentum carried by the density wave to the disk fluid at the rate $-\partial_R F_{\rm J}$ per unit radial distance, where $F_{\rm J}$ is the angular momentum flux carried by the wave. 

In shock-mediated disks, the angular momentum deposition rate $\partial_R F_{\rm J}$ replaces the stress term in the right hand side of equation (\ref{eq:AMcons}), so that the angular momentum evolution is governed by 
\ba
R^2\partial_t\langle\Sigma \delta v_\phi\rangle-\dot M\partial_R l = \partial_R F_{\rm J}.
\label{eq:AMconsShock}
\ea
As a result, the mass accretion rate in a shock-mediated disk can be expressed as
\ba   
\dot M = -\left(\partial_R l\right)^{-1}\left(\partial_R F_{\rm J} - R^2\partial_t\langle\Sigma \delta v_\phi\rangle\right).
\label{eq:MdotGen}
\ea   
This expression generalizes a similar result of \citet{Rafikov16}, who neglected the last time-dependent term in equation (\ref{eq:MdotGen}) finding  \citep{lynden-bell_1974}
\ba
\dot M=-(\partial_R l)^{-1}\partial_R F_{\rm J}.
\label{eq:simple_Mdot}
\ea
This simplification is legitimate when the disk is in steady state, which is generally not true ( in particular, in our work, see \S \ref{sect:time-dep}). For that reason, here we will use a more complicated expression (\ref{eq:MdotGen}). 

\citet{Rafikov16} demonstrated theoretically that the angular momentum deposition rate $\partial_R F_{\rm J}$ due to damping of a spiral shock with azimuthal wavenumber $m$ and pattern speed $\Omega_{\rm p}$ is given by 
\ba
\partial_R F_{\rm J}=\partial_R F_{\rm J}^{\rm th} = \mbox{sgn}[\Omega_{\rm p}-\Omega(R)]~mR\Sigma_{\rm ps} c_{\rm ps}^2~\psi_Q(\Pi),
\label{eq:dFJdr}
\ea
where $c_{\rm ps}(R)$ is the {\it pre-shock} value of the isothermal sound speed $c_{\rm s}$ at the radius $R$, and $\Sigma_{\rm ps}$ is the pre-shock surface density.  Also, an auxiliary function $\psi_Q(\Pi)$ of the ratio $\Pi\equiv p/p_{\rm ps}$ of the post-shock ($p$) to pre-shock ($p_{\rm ps}$) pressure is given by 
\ba
\psi_Q(\Pi) &\equiv & \frac{1}{(\gamma-1)}\left[\Pi\left(\frac{\gamma+1+(\gamma-1)\Pi}{\gamma-1+(\gamma+1)\Pi}\right)^{\gamma}-1\right]
\label{eq:psiQ}
\ea  
for an ideal gas with polytropic index $\gamma>1$. In practical applications one can often approximately set $\Sigma_{\rm ps}$ and $c_{\rm ps}$ in equation (\ref{eq:dFJdr}) to their mean (i.e. azimuthally averaged) values at a given radius. 

Another convenient way to express irreversible heating is via the density contrast across the shock $\rho/\rho_{\rm ps}$ ($\rho$ and $\rho_{\rm ps}$ being the post- and pre-shock values of gas density):
\ba
\psi_Q = \frac{1}{(\gamma-1)}\left[\left(\frac{\rho_{\rm ps}}{\rho}\right)^{\gamma}
\frac{(\gamma+1)(\rho/\rho_{\rm ps})-(\gamma-1)}{(\gamma+1)-(\gamma-1)(\rho/\rho_{\rm ps})}-1\right].
\label{eq:psiQ_rho}
\ea 
Note that in our 2D treatment of the disk $\rho/\rho_{\rm ps}$ should be replaced with $\Sigma/\Sigma_{\rm ps}$.

In the case of an isothermal equation of state ($\gamma=1$) one finds from expression (\ref{eq:psiQ}), taking the limit $\gamma\to 1$, that
\ba
\psi_Q &=& \frac{\epsilon(2+\epsilon)-2(1+\epsilon)\ln(1+\epsilon)}{2(1+\epsilon)},
\label{eq:isoth}
\ea
where $\epsilon \equiv (\Sigma-\Sigma_{\rm ps})/\Sigma_{\rm ps}=(p-p_{\rm ps})/p_{\rm ps}$ is the relative density (or pressure) jump across the shock. This expression is valid for arbitrary amplitude of $\epsilon$, and agrees with similar result found in \citet{Belyaev2}.

At small wave amplitudes ($\Delta\Sigma/\Sigma_{\rm ps}\ll 1$) all these expressions reduce to the well-known cubic scaling \citep{Spruit87,Larson,Savonije}
\ba
\psi_Q \approx \frac{\gamma(\gamma+1)}{12}\left(\frac{\Delta\Sigma}{\Sigma_{\rm ps}}\right)^3.
\label{eq:cubic}
\ea
However, at high wave amplitudes this approximation becomes inaccurate, as we show later (see \S \ref{sect:ampl}).

The main goal of this work is to numerically verify the validity of equations (\ref{eq:MdotGen}), (\ref{eq:dFJdr})-(\ref{eq:isoth}). We do this in \S  \ref{sect:results}-\ref{sect:time-dep}.


\section{Numerical setup}  
\label{sect:num_setup}


We run a set of two-dimensional numerical simulations in cylindrical polar  coordinates $(R,\phi)$ using publicly available code Athena++\footnote{\url{http://princetonuniversity.github.io/athena/}} (Stone et al. 2017, in preparation), the grid-based code, which solves equations of gas dynamics using high-order Godunov methods. This code is a new version of a popular astrophysical gas dynamics code Athena \citep{Athena1,Athena2}, and features improved performance and scalability as well as more flexible choice of coordinate systems. For all our simulations we use HLLE solver \citep{1988SJNA...25..294E}.
Athena++ was extensively used in simulations of protoplanetary disks \citep{Dong_spiral,Zhu,ArzZhuSt2017} and cataclysmic variables \citep{Ju,Ju2017}, and was proven to accurately reproduce the properties of density waves. To provide a direct comparison with the analytical predictions of  \citet{Rafikov16}, we consider a gaseous disk with no explicit viscosity (we describe the effects of numerical dissipation in \S \ref{sect:res}).
 
The active domain of our simulations $(R,\phi) \in [0.4,1.0]\times[0,\pi]$, i.e. it is limited to only half the disk in $\phi$ direction. We do that to increase resolution for a given computational cost. As we show in \S\ref{sect:res}, high resolution is crucial in order to demonstrate agreement between theory and simulations. Our typical resolution is 447 cells per scale-height at the outer radius $R_0$ (or $447/H_0$ for brevity). 

We typically run our simulations for $20 P_0$, where $P_0=2\pi/\Omega_0$ is the orbital period at the outer radius. This is sufficient for the global wave pattern to fully develop inside the domain and reach a steady state. However, this not long enough for the disk material to get radially redistributed by shocks for $\dot M=$ const to be achieved, i.e. the disk itself does not relax to a global equilibrium (cf. \citealt{Ryan}). For that reason the first, time-dependent term in equations (\ref{eq:AMcons})-(\ref{eq:AMconsShock}) plays important role in the angular momentum balance. This has important consequences discussed in \S \ref{sect:time-dep}.

Our strategy consists of running a number of simulations, in which only one parameter is varied as compared to a certain fiducial run. This allows us to get a clear picture of the effect of different physical inputs on the results. The numerical parameters used in our fiducial run are listed in Table \ref{tab:parameter}. 

\begin{figure}
\centering
\includegraphics[width=0.47\textwidth]{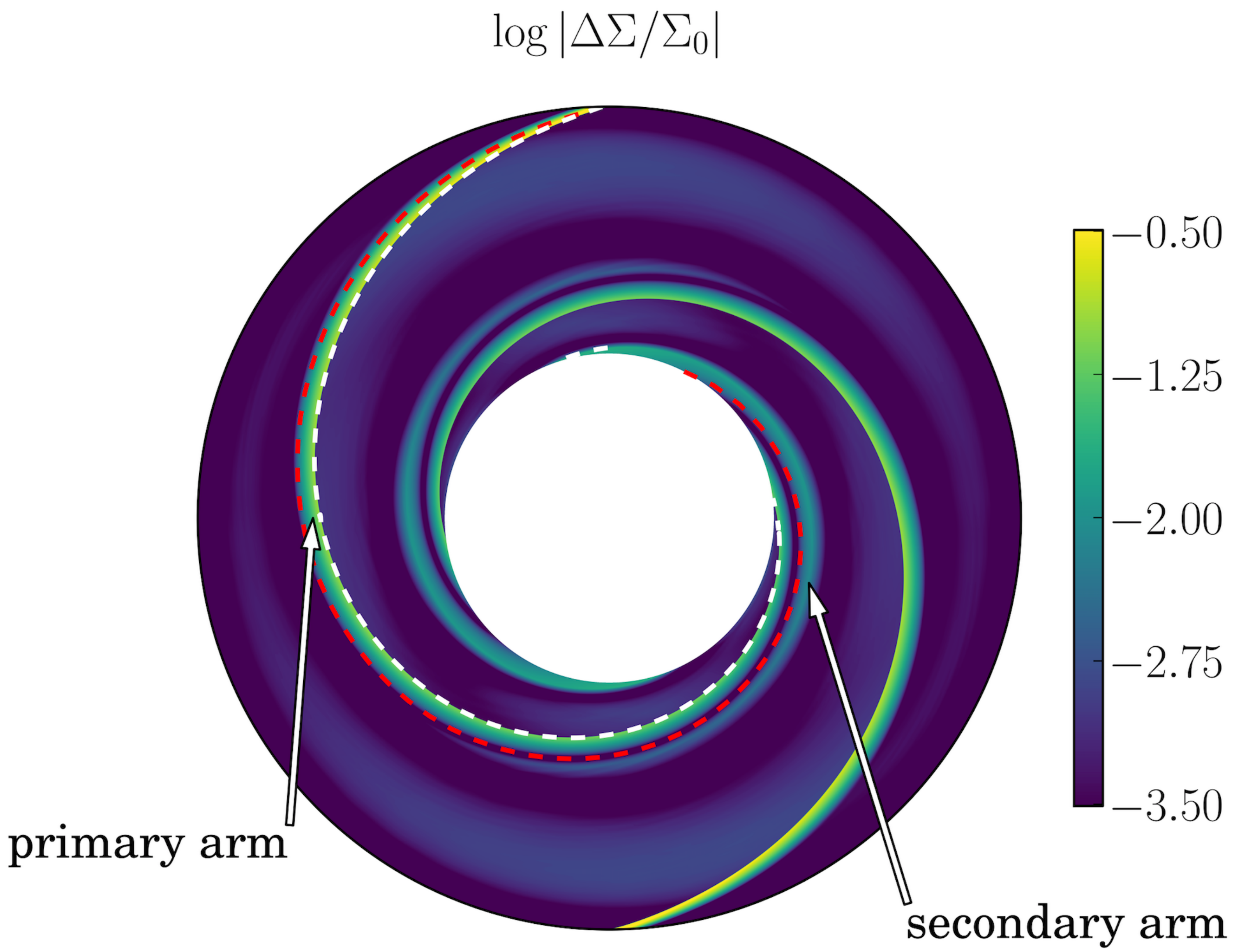}
\caption{Two-dimensional map of the (logarithm of the) surface density perturbation emerging in one of our typical simulations at $t=10P_0$
($P_0=2\pi/\Omega_0$ is the orbital period at the outer radius $R_0=1$) having an amplitude of $A_0=1$ at the outer boundary of the computational domain. This simulation features a fiducial set of parameters, namely a resolution of $447/H_0$, aspect ratio $H_0/R_0=0.2$ at the outer radius, radially constant background surface density and temperature profiles, and the isothermal equation of state for the gas (see Table \ref{tab:parameter}). The red dashed curve indicates the location of the spiral wake predicted by the linear (low-amplitude) theory \citep{Rafikov02,Ogilvie}. The actual location of the wave crest (coincident with the shock position) is shown by the white dashed curve. It deviates from the linear theory prediction because of the non-linear effects \citep{Rafikov02,Zhu}. 
\label{fig:density_map}}
\end{figure}


\subsection{Initial conditions}
\label{sect:ICs}


We initialize the disk with the equilibrium model described in \citet{Nelson2013}, which has power-law profiles of the surface density and temperature:
\begin{align}
\Sigma_{\rm init}(R) &= \Sigma_0 (R/R_0)^p,\label{eq:Sprofile}\\
T_{\rm init}(R) &= T_0 (R/R_0)^q,\label{eq:Tprofile}
\end{align}
where $\Sigma_0$ and $T_0$ correspond to the outer radius $R_0$ of the simulation domain. The hydrodynamic equilibrium implies that the angular velocity of the disk
\begin{equation}
    \Omega^2(R) = \Omega_{\rm K}^2(R) \left[1 + (p+q)\left(\frac{c_{\rm s}/\Omega_{\rm K}}{R}\right)^2 \right],
    \label{eq:Omega}
\end{equation}
where $\Omega_{\rm K}(R) = (G M_\star/R^3)^{1/2}$ is the Keplerian frequency, and $c_{\rm s}(R) = c_0 (R/R_0)^{q/2}$ is the local isothermal sound speed. It is customary to describe the temperature of the disk using the aspect ratio:
\begin{equation}
  H(R)/R = H_{\rm 0}/R_{\rm 0} (R/R_{\rm 0})^{(q+1)/2}.
\end{equation}

For our fiducial run we use the uniform density profile with $p = 0$, as well as the globally isothermal equation of state $\gamma = 1$, $q = 0$. The non-uniform density profiles are considered in \S \ref{sect:Sigma}, and we study other equations of state in \S \ref{sect:EOS}. We adopt $H_0/R_0 = 0.2$ as our fiducial disk aspect ratio. We modify the aspect ratio and study its effect on the results in \S \ref{sect:h-r}.

\begin{table}[]
\small
\caption{Simulation set up\label{tab:parameter}}
\begin{tabular}{ll}
Parameters & Values\\
\hline
\hline
Domain $[R]\times [\phi] $ & $[0.4, 1.0]\times [0,\pi]$\\
$N_R\times N_\phi$ & $2048 \times 7168$\\
$H_0/R_0$ & 0.2\\
Resolution & $283/H_{\rm in} = 447/H_0 = 91/\Delta\phi$
 \end{tabular}
\end{table}


\subsection{Boundary conditions and wave triggering}
\label{sect:BCs}


To launch the density wave we impose a boundary condition in the form of a density perturbation at the outer radius of the disk. The shape of the perturbation in $\phi$ sets the initial profile of the density wave, and is given by
\ba
   \delta \Sigma_{\rm out}(\phi,t) & \propto & |\sin(\phi-\phi_0(t))|^{1/2}
   \nonumber\\
   & \times &
   \exp\left\{-\left[\frac{\phi-\phi_0(t)}{\Delta\phi}\right]^2\right\},
   \label{eq:SigPert}
\ea
where $\phi_0(t) = \Omega_{\rm p} t$ sets the rotation of the perturbation pattern at angular speed $\Omega_{\rm p}$, and $\Delta \phi$ describes the initial width of the profile. We use $\Delta \phi = 0.04$ for all of the simulations (this azimuthal range is resolved by about 90 grid points, see Table \ref{tab:parameter}). The pattern speed is usually chosen to be $\Omega_{\rm p} = 0$ for simplicity. We study the effects of nonzero $\Omega_{\rm p}$ in \S \ref{sect:rot}.

We set the velocity perturbation in $R$ and $\phi$ directions to zero at the outer boundary. At the inner disk boundary we reset all the quantities to their initial values.

In our simulations we vary the proportionality coefficient in equation (\ref{eq:SigPert}) to have different perturbation amplitude $A_0={\rm max}\left[\delta \Sigma_{\rm out}/\Sigma(R_0)\right]$. As the simulation evolves, the density perturbation propagates into the active domain in the form of a spiral density wave, which rapidly turns into a shock. The steep profile $\propto |\sin(\phi-\phi_0)|^{1/2}$ for $|\phi - \phi_0| \ll \Delta\phi$ assures that the shock develops very close to the outer boundary.

We do not expect our results to be affected by the details of the implementation of our boundary conditions, as long as the wave shocks close to the outer boundary. However, some fine features, like the appearance of a secondary shock (see \S \ref{sect:nontrivial}), may be more sensitive to the boundary conditions. Careful examination of this issue is deferred to future study.

Since we simulate only one half of the disk, we use periodic boundary conditions to connect the values of fluid variables at $\phi=0$ and $\pi$. As a result, the developing perturbation pattern features two identical density waves separated by $\pi$ in azimuth, see Figure \ref{fig:density_map}. When computing angular momentum contributions using equation (\ref{eq:AMcons}), we account for both shocks.

\begin{figure}
\centering
\includegraphics[width=0.5\textwidth]{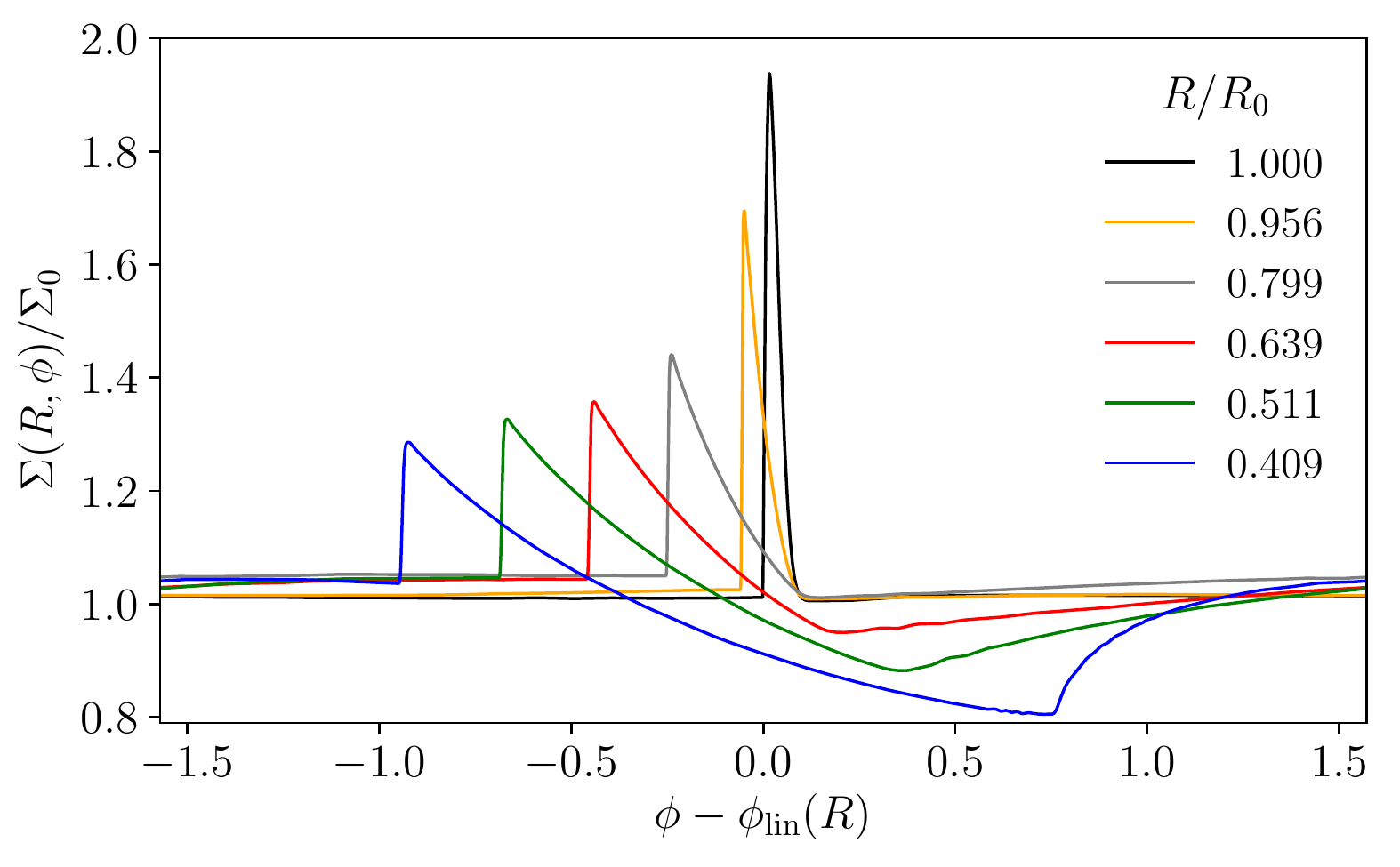}
\caption{Evolution of the profile of the density perturbation shown in Figure \ref{fig:density_map}. Azimuthal cuts through the spiral wake at different radii are shown, referenced to the azimuthal angle $\phi_{\rm lin}(R)$ corresponding to the linear prediction for the density wake location \citep{Rafikov02,Ogilvie}. One can see that nonlinear evolution and damping cause continuous distortion of the wave profile, making it broader and reducing its amplitude.
\label{fig:profile}}
\end{figure}


\subsection{Shock detection}
\label{sect:shock}


To calculate the density jumps across the shocks we employ a simple algorithm to detect their location. At each radius, we calculate $\partial_\phi \log \Sigma(R,\phi)$ and identify the shock with the azimuthal region within which this derivative is greater than 1. Having done that, we calculate the number of shocks, the jump conditions across the shock, as well as the azimuthal width of the wake.

As shock has very steep density profile, the location of the shock is not very sensitive to the details of the detection algorithm\footnote{Although some other aspects of our calculations may be sensitive to it, see \S \ref{sect:disc}.}. We have verified by eye that our algorithm correctly reproduces both the location and the amplitude of the shock at all radial positions. The shock profile in $R-\phi$ coordinates resulting from this algorithm is illustrated in Figure \ref{fig:density_map}.


\section{Results}  
\label{sect:results}


We now provide a brief description of certain characteristic features of our simulations, before giving a more in-depth, systematic analysis of their results in \S \ref{sect:theory}.

Figure \ref{fig:density_map} is a 2D color map of the surface density perturbation in one of our simulations with the fiducial set of parameters listed in Table \ref{tab:parameter}. This particular run features surface density perturbation with the amplitude of $A_0 = 1$ at the outer edge of the simulation domain, see \S \ref{sect:num_setup}. One can see that the perturbation propagates through the radial extent of the disk in the form of a coherent spiral wave all the way to the inner boundary. In our runs we do not observe the development of the spiral wave instability, previously reported by \citet{Kim}, \citet{Bae}, and \citet{Sormani}. It is not clear whether this outcome is caused by the lack of an external gravitational perturber in our simulations, or their relatively limited radial extent.

\begin{figure}
\centering
\includegraphics[width=0.5\textwidth]{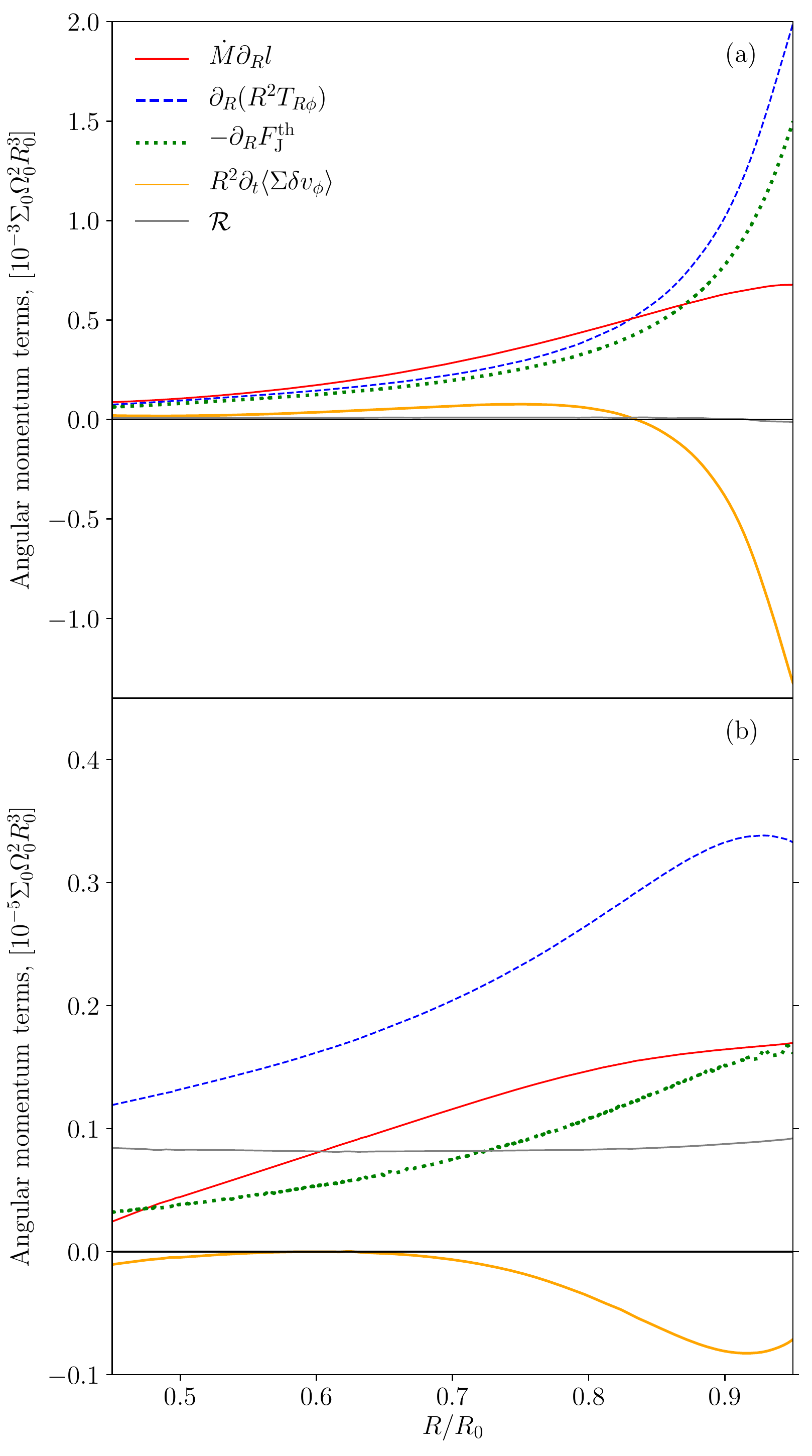}
\caption{Radial profiles of the different angular momentum contributions appearing in equations (\ref{eq:AMconsShock}) and (\ref{eq:residual}), obtained from simulations with different initial wave amplitudes at the outer boundary: (a) $A_0=1$ and (b) $A_0=0.0625$. Red solid, blue dashed and green dotted lines show the advective term $\dot M\partial_R l$, stress term $\partial_R (R^2 T_{R\phi})$, and theoretical prediction $\partial_R F_{\rm J}^{\rm th}$ for the stress contribution, correspondingly. The latter is computed via equations (\ref{eq:dFJdr})-(\ref{eq:isoth}) using the density jump at the shock measured from simulations. The orange curve shows the time-dependent term $R^2\partial_t\langle\Sigma \delta v_\phi\rangle$, while the grey curve is the residual term ${\cal R}$, see equation (\ref{eq:residual}).
\label{fig:AM_6.25}}
\end{figure}

\begin{figure*}
\centering
\includegraphics[width=0.5\textwidth]{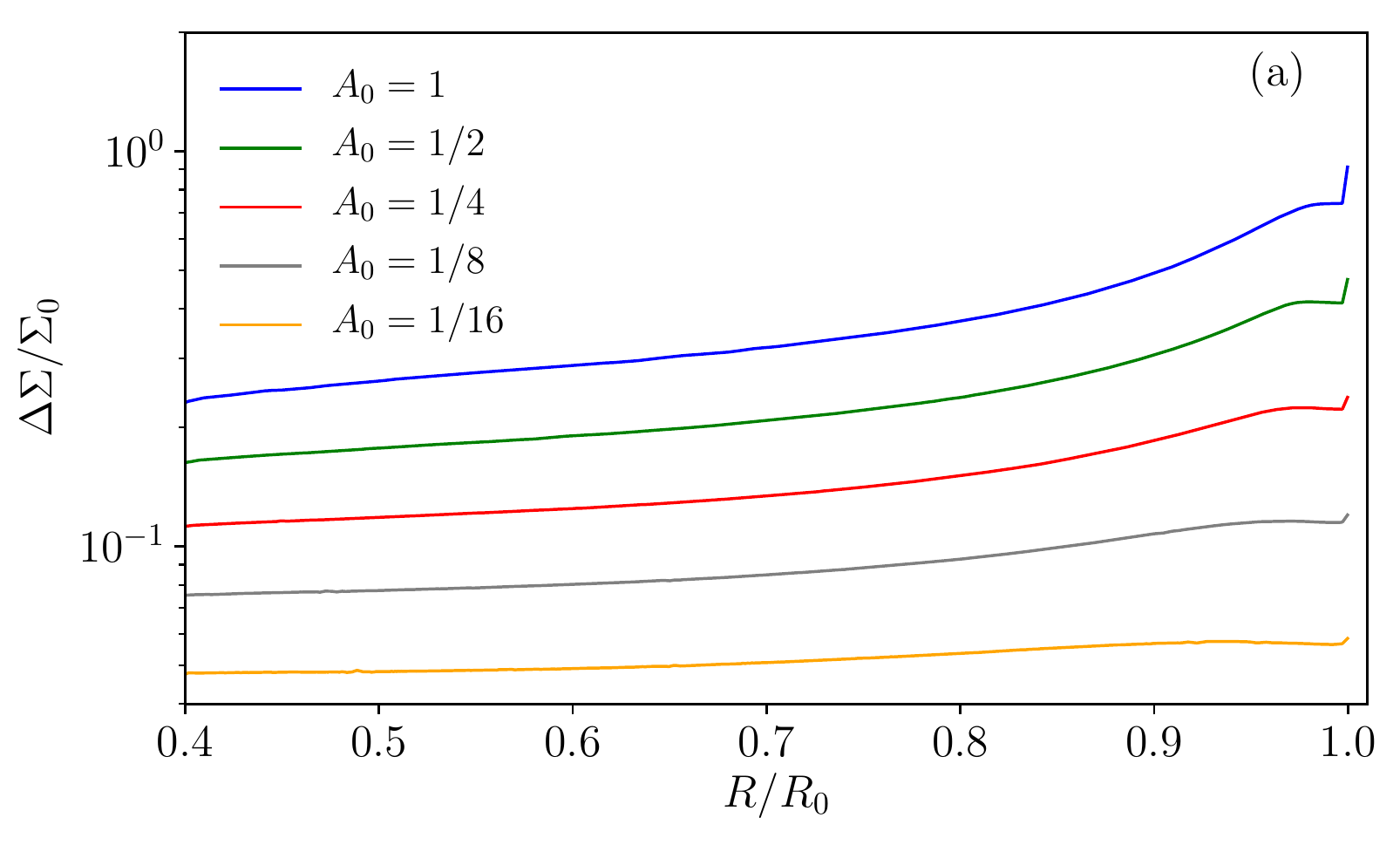}~\includegraphics[width=0.5\textwidth]{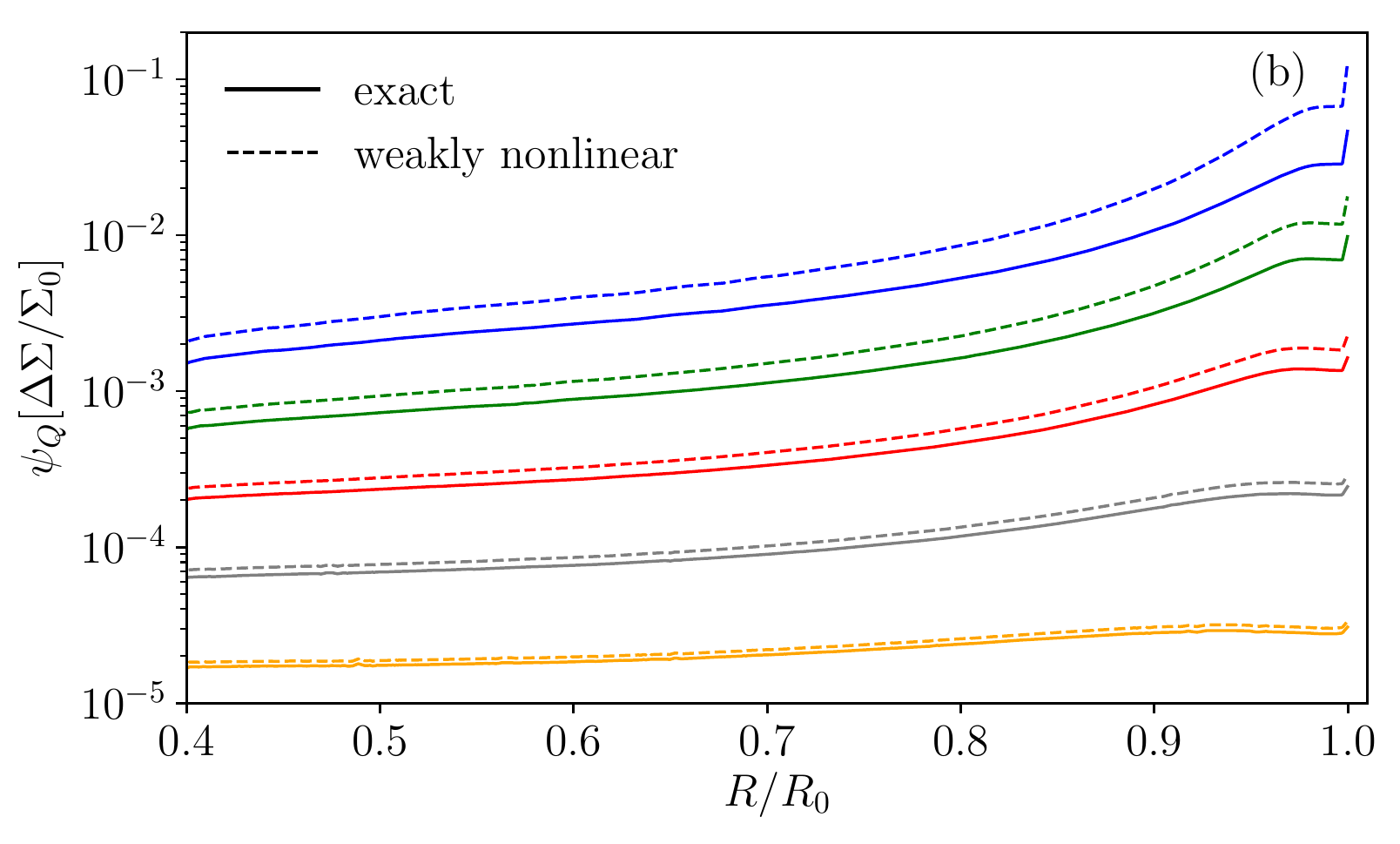}
\caption{(a) Radial profiles of the density jump across the shock measured in simulations with different values of the initial wave amplitude $A_0$ (indicated in panel).  (b) Radial profiles of the auxiliary function $\psi_Q(\Delta\Sigma/\Sigma)$ (solid line), given by equation (\ref{eq:isoth}) for an isothermal equation of state and density jumps from the panel (a). Dashed lines show the weakly nonlinear approximation given by equation (\ref{eq:cubic}). This approximation clearly overpredicts the effect of the shock on the disk, especially at high wave amplitudes. 
\label{fig:psiQ}}
\end{figure*}

The red dashed curve in Figure \ref{fig:density_map} shows the {\it linear} prediction for the shape of the spiral wake based on the calculation in \citet{Rafikov02} [see equation (36) of that work]. One can see that the actual location of the wave crest deviates from the linear approximation in such a way as to make the spiral less tightly wound in the central part of the disk than the linear theory would predict. This deviation is caused by significant nonlinear distortion and broadening of the azimuthal profile of a high amplitude wave \citep{Rafikov02}. Radial profile of the surface density jump across the shock relative to the local background density $\Delta\Sigma/\Sigma_0$, characterizing the evolution of the wave nonlinearity, is shown by the blue curve in Figure \ref{fig:psiQ}a, which is discussed further in \S \ref{sect:ampl}.

\begin{figure*}
\centering
\includegraphics[width=1\textwidth]{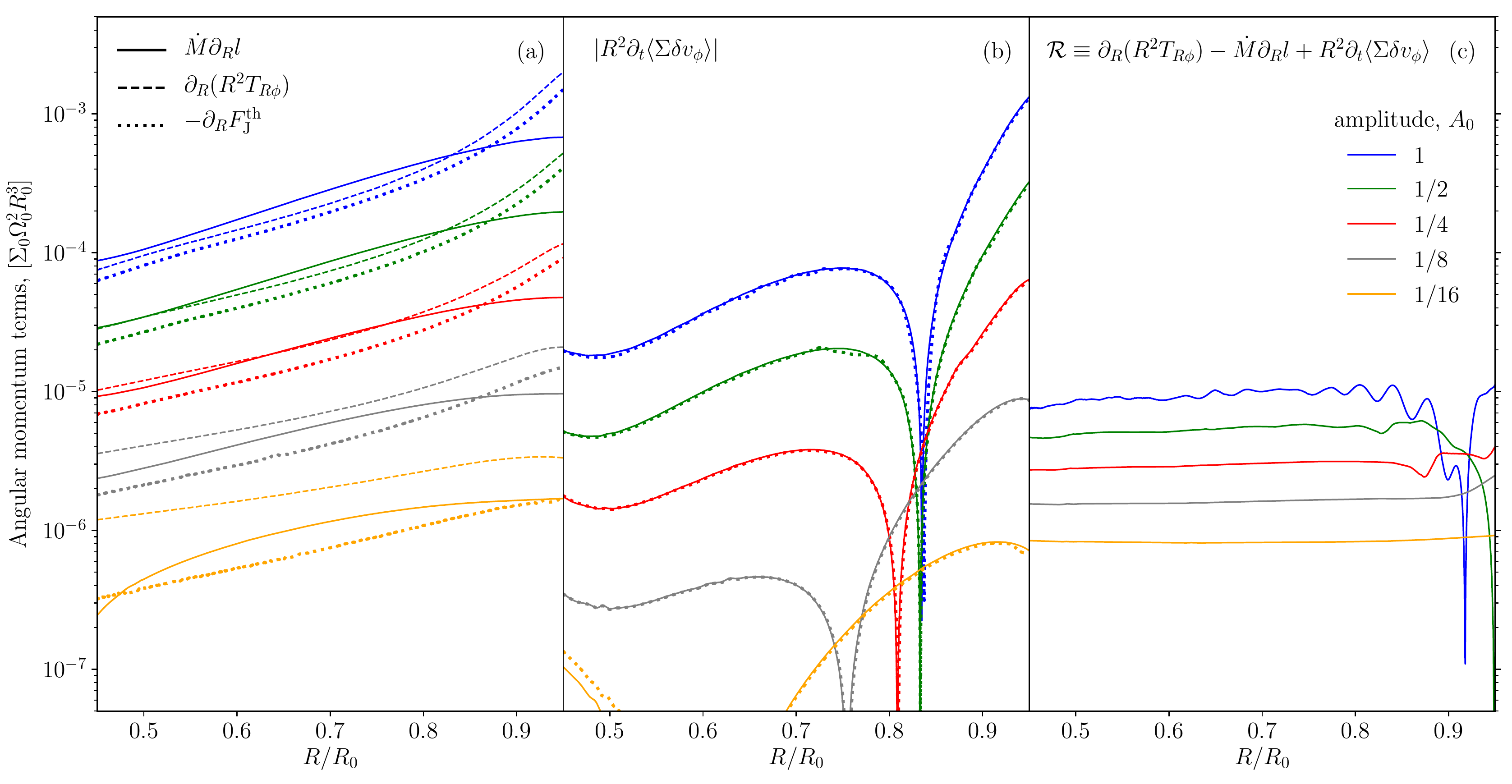}
\caption{Radial profiles of the angular momentum contributions appearing in equations (\ref{eq:AMconsShock}) and (\ref{eq:residual}), obtained from simulations with different initial wave amplitude at the outer boundary $A_0$ (shown by different colors, see legend in the right panel). (a) Solid, dashed and dotted lines show the advective term $\dot M\partial_R l$, stress term $\partial_R\left(R^2 T_{R\phi}\right)$, and theoretical prediction $\partial_R F_{\rm J}^{\rm th}$ for the stress contribution, correspondingly (as in Figure \ref{fig:AM_6.25} but on log vertical scale).  (b) The absolute value of the time-dependent term $R^2\partial_t\langle\Sigma \delta v_\phi\rangle$ (solid curves). Colored dots of corresponding color show theoretical prediction for the time-dependent term computed using equation (\ref{eq:time_dep}), which is in excellent agreement with the numerical results. See \S \ref{sect:time-dep} for more details. (c) The residual term ${\cal R}$ defined by equation (\ref{eq:residual}), which is caused by numerical dissipation in our simulations.
\label{fig:AMterms}}
\end{figure*}

The evolution of the azimuthal profile of the wake is illustrated in Figure \ref{fig:profile}, which displays azimuthal cuts (at several radii) of the surface density (normalized by the density at the outer boundary) around the azimuthal angle $\phi_{\rm lin}(R)$ corresponding to the linear prediction for the wake position. One can clearly see that, as a result of the nonlinear effects, the wake shape gets continuously distorted, resulting in the decay of its amplitude and profile broadening as the wave propagates towards the disk center. The importance of properly accounting for the wave non-linearity (as opposed to using purely linear results) for explaining the openness of the spiral arms driven by massive perturbers in the protoplanetary disks has been previously emphasized in \citet{Dong_spiral} and \citet{Zhu}. 

As the main focus of our paper is on verifying the analytical predictions of \citet{Rafikov16} for the shock-driven mass and angular momentum transport, we extract from our simulations and analyze a number of relevant quantities. In particular, Figure \ref{fig:AM_6.25} illustrates the radial profiles of the different angular momentum contributions appearing in the equation (\ref{eq:AMcons}), directly measured in the simulation with the fiducial disk parameters and the perturbation amplitudes $A_0$ of 1 and 0.0625 at the outer radius. In the order that they appear in this equation, we call these contributions the {\it time-dependent} term (as it involves the time-derivative of disk variables) $R^2\partial_t\langle\Sigma \delta v_\phi\rangle$, the {\it advective} term $\dot M\partial_R l$, and the {\it stress} term $\partial_R\left(R^2 T_{R\phi}\right)$. To ensure better statistics, all terms are averaged over $5P_0$ after the wave pattern fully develops. It is clear that the radial profiles of the advective and stress terms differ in shape, which is caused by the substantial contribution coming from the time-dependent term. The difference in especially pronounced close to the outer disk radius, where the time-dependent term is most significant.

One can also see that the three terms do not sum up to zero, as they should according to the equation (\ref{eq:AMcons}). The remaining non-zero residual (thin grey line) is explicitly defined as 
\ba
{\cal R}\equiv R^2\partial_t\langle\Sigma \delta v_\phi\rangle-\dot M\partial_R l +\partial_R\left(R^2 T_{R\phi}\right),
\label{eq:residual}
\ea   
see equation (\ref{eq:AMconsShock}). It is roughly independent of radius and in simulation with $A_0=0.0625$ has a magnitude comparable to other angular momentum contributions. The fact that ${\cal R}$ is not equal to zero is caused by the numerical dissipation intrinsic for our finite-resolution runs; this is discussed in more details in \S \ref{sect:ampl} and \ref{sect:res}.

In Figure \ref{fig:AM_6.25} we also show the analytical prediction $\partial_R F_{\rm J}^{\rm th}$ (with negative sign, so it can be directly compared with advective and stress terms) given by the equation (\ref{eq:dFJdr}) and calculated using the value of the density jump at the shock front measured in our simulations. Figure \ref{fig:AM_6.25}b reveals that  $\partial_R F_{\rm J}^{\rm th}$ has a radial profile similar to the stress term, but with a vertical offset approximately equal to the numerical residual term ${\cal R}$.  

Figure \ref{fig:AM_6.25}a is identical to Figure \ref{fig:AM_6.25}b, except that it corresponds to a run with a higher amplitude of the perturbation at the outer edge, $A_0=1$. Stronger perturbation clearly results in a significant increase in the amplitude of all physical angular momentum contributions, compared to Figure \ref{fig:AM_6.25}b. At the same time, the numerical residual does not increase nearly as much, rendering its contribution to the angular momentum balance (i.e. equation (\ref{eq:AMcons})) insignificant. This results in a considerably better agreement between the theoretical $\partial_R F_{\rm J}^{\rm th}$ and the stress term at high wave amplitudes. Also, the time-dependent term is still important and leads to a significant difference between the profiles of the advective term and both the stress term and the theoretical prediction. We provide a more systematic discussion of the role of the amplitude $A_0$ on the wave propagation and associated transport properties in \S \ref{sect:ampl}.


\section{Comparison with theory}  
\label{sect:theory}


We now provide a more detailed comparison between the theory presented in \citet{Rafikov16} and our hydro simulations. The theory predicts that the effect of shocks on the disk depends on many physical parameters: strength of the shock, aspect ratio of the disk, radial profiles of the background density and temperature, equation of state, and so on. Our goal is to thoroughly test these dependencies. In addition, we will explore the sensitivity of our results to numerical parameters, such as the resolution of our simulations. This will help us formulate the conditions, under which one could expect simulations to reliably reproduce the actual effect of the global spiral shocks on disk evolution.


\subsection{Variation of the wave amplitude.}  
\label{sect:ampl}


We start by exploring the effect of the wave amplitude on the correspondence between the theory and simulations.  In Figure \ref{fig:AMterms} we plot different angular momentum contributions appearing in equation (\ref{eq:AMcons}) as a function of radius, for different values of the wave amplitude at the outer radius. 

In the left panel of this Figure we plot the stress term $\partial_R\left(R^2 T_{R\phi}\right)$ together with the theoretical prediction $\partial_R F_{\rm J}^{\rm th}$, as well as the advective term $\dot M\partial_R l$. One can see a very rapid increase in the magnitude of all angular momentum contributions with the amplitude $A_0$ of the initial surface density perturbation. The top and bottom sets of curves correspond to the values of $A_0$ different by a factor of 16, but the associated angular momentum contributions differ by more than three orders of magnitude near the outer edge.  

Similar to the pattern described in \S \ref{sect:results}, we see  good agreement between the $\partial_R\left(R^2 T_{R\phi}\right)$ and $\partial_R F_{\rm J}^{\rm th}$. Not only does the theory correctly predict the variation of the shock-driven stress with $A_0$, the two curves also closely follow each other in radius, with just a small vertical offset separating them. The offset becomes less pronounced for larger $A_0$. Note that theoretical curve always underpredicts the numerical result, but only at the level of several tens of per cent or less at high wave amplitudes. The reasons for this discrepancy are discussed in \S \ref{sect:disc}.

In the linear regime ($A_0\ll 1$) the agreement between $\partial_R\left(R^2 T_{R\phi}\right)$ and $\partial_R F_{\rm J}^{\rm th}$ is much worse, as already mentioned in \S \ref{sect:results}. To illustrate the reasons, in the right panel of Figure \ref{fig:AMterms} we display the variation of the residual ${\cal R}$ with the wave amplitude. One can see that at the lowest amplitudes ${\cal R}$ is comparable to the stress and advective terms in magnitude. However, as $A_0$ increases, the residual term grows much slower than the physical angular momentum contributions. As a result, the quantitative agreement between $\partial_R\left(R^2 T_{R\phi}\right)$ and $\partial_R F_{\rm J}^{\rm th}$ is much better for more nonlinear waves (higher $A_0$). We discuss the origin and behavior of ${\cal R}$ further in \S \ref{sect:res}.

Similar to Figure \ref{fig:AM_6.25}, the left panel of Figure  \ref{fig:AMterms} also shows a significant discrepancy between the advective and the stress terms. The qualitative difference is present at all wave amplitudes, and is especially pronounced in the outer disk. As in \S \ref{sect:results}, this difference is naturally explained for all values of $A_0$ by the large time-dependent contribution $R^2\partial_t\langle\Sigma \delta v_\phi\rangle$. Its absolute value is shown in the middle panel of Figure \ref{fig:AMterms}. It is comparable to the stress and advective contributions in the outer disk, showing the importance of accounting for this angular momentum term in general. This implies that in our calculations we cannot accurately predict the radial behavior of $\dot M$ just by using the steady state equation (\ref{eq:simple_Mdot}) and the shock dissipation prescription (\ref{eq:dFJdr}). Determination of the accretion rate throughout the disk in general requires the use of full equation (\ref{eq:MdotGen}), properly accounting for the time-dependent term, which is discussed further in \S \ref{sect:time-dep}.

In Figure \ref{fig:psiQ}a we display the radial profiles of the relative surface density perturbation  $\Delta\Sigma/\Sigma$ for different $A_0$. One can see a clear variation in the behavior as the wave nonlinearity changes. At the lowest amplitudes $\Delta\Sigma/\Sigma$ remains roughly constant as the wave travels inwards. However, at high values of $A_0$ density perturbation rapidly decays with $R$: for $A_0=1$ the value of $\Delta\Sigma/\Sigma$ drops by about a factor of 3 as the wave propagates from $R_0$ to $0.4R_0$. This decay of the perturbation is caused by the much stronger nonlinear wave damping at higher values of $A_0$. As a result, the angular momentum terms in Figure \ref{fig:AMterms}a also show much faster inward decay for higher $A_0$. 

\begin{figure*}
\centering
\includegraphics[width=1\textwidth]{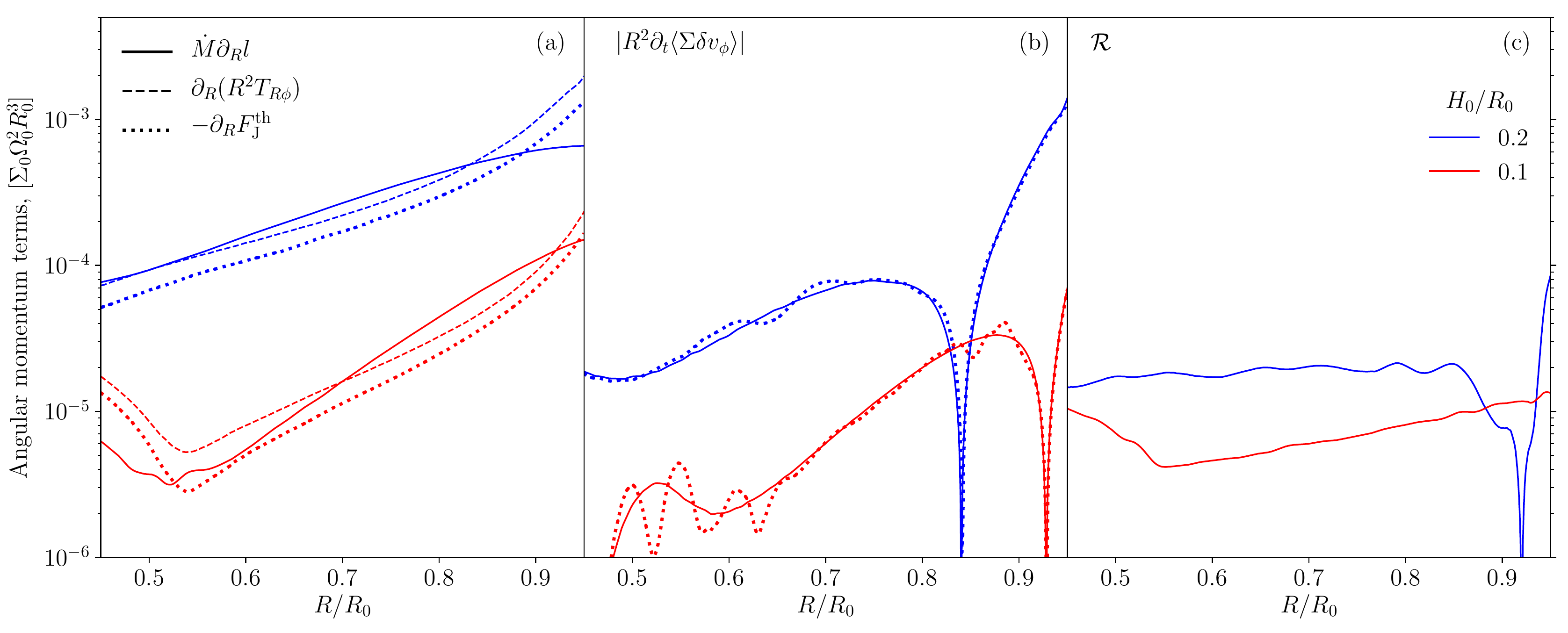}
\caption{
Same as Figure \ref{fig:AMterms} but now illustrating the effect of changing disk scale height $H_0/R_0$ (values indicated in panel (c)) on the angular momentum flux contributions. Resolution per scale height $H_0$ is the same ($223/H_0$) for these two simulations. Angular momentum deposition rate by the shock shows faster decay with the distance travelled from the outer edge in a colder disk. Note the rise of the shock-induced stress in the inner part of the disk with $H_0/R_0=0.1$, explained by the appearance of a secondary shock, see \S \ref{sect:nontrivial}.
\label{fig:hor}}
\end{figure*}

We also note a very sharp drop in the amplitude of the surface density perturbation at the outer boundary of the disk. This artifact of our boundary conditions arises because we do not initialize velocity perturbation consistently with density perturbation at the boundary. As a result, the actual amplitude of the wave close to the boundary is slightly different from $A_0$. However, this does not affect the comparison between theory and simulations as $\partial_R F_{\rm J}^{\rm th}$ is always computed using the actual local value of the surface density perturbation.

Figure \ref{fig:psiQ}b illustrates how the simple cubic approximation (\ref{eq:cubic}) describes the effect of the shock compared to the fully nonlinear prescription (\ref{eq:psiQ})-(\ref{eq:isoth}); we plot the full $\psi_Q$ given in the isothermal case by the equation (\ref{eq:isoth}) as well as the small amplitude approximation (\ref{eq:cubic}). The two clearly agree with each other quite well at the low values of $A_0$. However, as the  wave nonlinearity increases, the cubic approximation considerably overpredicts the actual effect of the shock on the disk; the discrepancy is close to a factor of 2 for $A_0=1$.


\subsection{Sensitivity to $H/R$.}  
\label{sect:h-r}


Next we vary disk aspect ratio $H/R$ and explore the effect on agreement with theory. In Figure \ref{fig:hor} we compare different angular momentum contributions for two simulations with $A_0=1$, in which we vary $H_0/R_0$ from $0.2$ to $0.1$. In doing so we kept the resolution {\it per scale height} the same, meaning that we had to decrease by a factor of 2 the number of grid points in every dimension in the case $H_0/R_0=0.2$. 

Comparing different sets of curves in Figure \ref{fig:hor} corresponding to different $H/R$, one can make several observations. First, irrespective of the value of $H/R$, our simulations demonstrate good agreement between the theoretical ($\partial_R F_{\rm J}^{\rm th}$) and numerical ($\partial_R\left(R^2 T_{R\phi}\right)$) shock-induced stress terms. The advective term clearly deviates from both because of the time-dependent contribution shown in Figure \ref{fig:hor}b.

Second, reduction of the disk scale height by a factor of 2 results in a dramatic reduction (by more than an order of magnitude at some radii) of all physical angular momentum contributions. Close to the outer edge of the disk this difference is consistent with our theoretical prediction (\ref{eq:dFJdr}) showing the explicit quadratic dependence of the shock-driven angular momentum flux on $c_{\rm s}$ (or, equivalently, $H/R$).

Third, the rate $\partial_R F_{\rm J}$ at which the density wave transfers angular momentum to the disk fluid drops faster as the wave propagates in colder disk. This is caused by the faster evolution of the wave for lower $H/R$ and, subsequently, more rapid decay of the angular momentum carried by the wave (also implying faster decay of $\Delta\Sigma/\Sigma$ across the shock). This effect is similar to the faster decay of the waves with higher initial amplitude, obvious in Figures \ref{fig:AMterms}a and \ref{fig:psiQ}a. For that reason, at radii of $R\approx 0.6 R_0$ the values of $\partial_R\left(R^2 T_{R\phi}\right)$ derived for two different $H/R$ differ more dramatically than near the outer edge of the disk. 

\begin{figure*}
\centering
\includegraphics[width=1\textwidth]{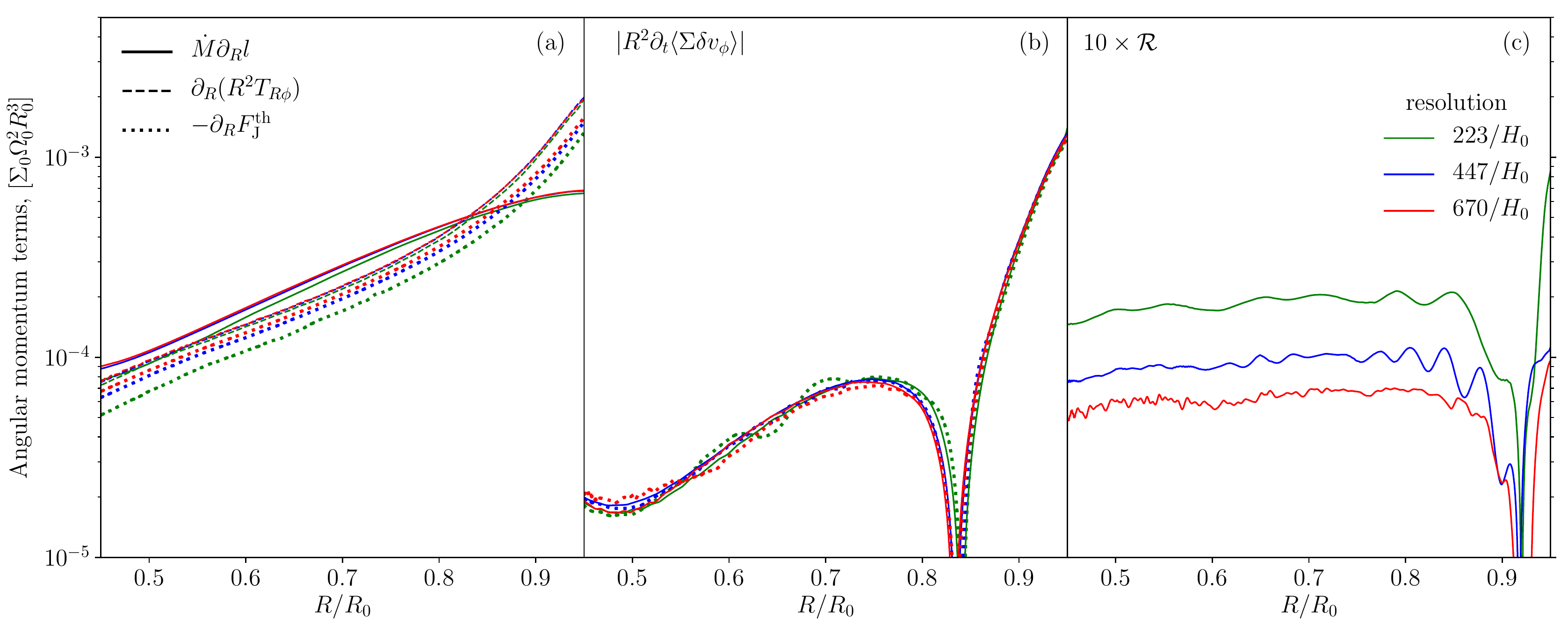}
\caption{
Same as Figure \ref{fig:AMterms} but now illustrating the effect of changing resolution on the agreement between the theory and numerics. 
\label{fig:resolution}}
\end{figure*}

Fourth, in colder disk the behavior of the stress term exhibits a qualitative change: its decay as the wave propagates inwards switches to {\it growth} at $R\lesssim 0.53R_0$. This change is caused by the faster evolution of the wave in a colder disk, resulting in emergence of a {\it secondary shock}. We will discuss this phenomenon in more detail in \S \ref{sect:nontrivial}, making only a couple of comments here. The new shock structure provides additional dissipation at its front, causing growth of $\partial_R\left(R^2 T_{R\phi}\right)$ as $R$ decreases (ultimately this new contribution to the stress term will also start to decay). In a hotter disk with $H_0/R_0=0.2$ we would find the same change in the behavior but at smaller radii (which are outside of our computational domain) because of the slower evolution of the wave. Nevertheless, regardless of the more complicated picture of the wave evolution in colder disks, our theory still reproduces  numerical results very well (apart from a small vertical offset).

\begin{figure}
\centering
\includegraphics[width=0.5\textwidth]{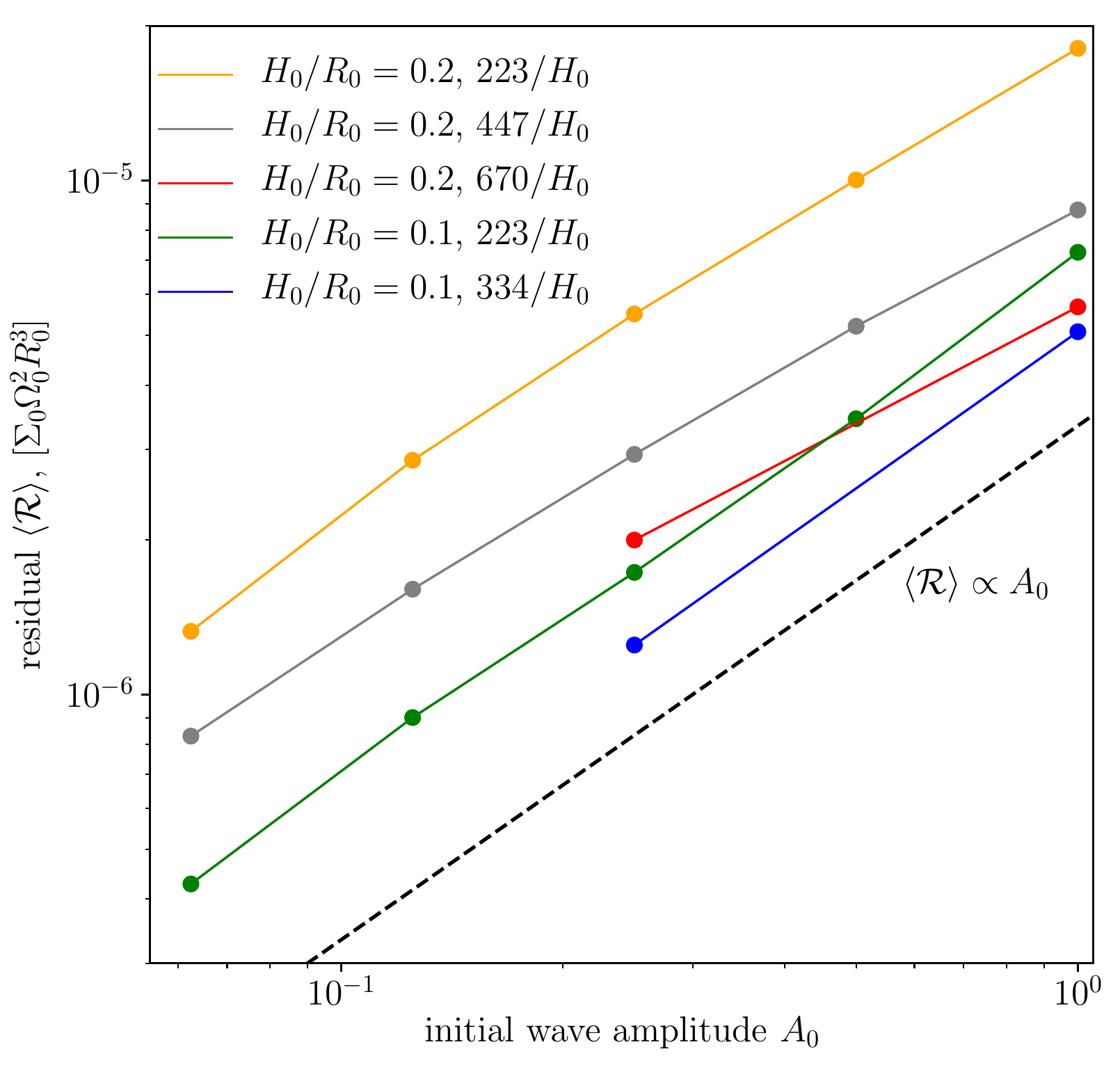}
\caption{
Residual term ${\cal R}$ plotted as a function of the wave non-linearity --- wave amplitude at the outer edge of the computation domain. Different curves correspond to different sets of the disk ($H_0/R_0$) and simulation (resolution per scale height $H_0$) parameters, indicated on the panel. Dashed line shows scaling ${\cal R}\propto A_0$. This figure illustrates that the residual goes down with both the disk aspect ratio $H_0/R_0$ and the numeical resolution.
\label{fig:residual}}
\end{figure}


\subsection{Effects of changing resolution.}  
\label{sect:res}


We now discuss the role of numerical effects on our results. In Figure \ref{fig:resolution} we display individual  contributions to the angular momentum budget obtained in simulations with three different resolutions. 

Figure \ref{fig:resolution}a shows that the shock-driven torque term $\partial_R (R^2 T_{R\phi})$ is insensitive to resolution. Its theoretical counterpart $\partial_R F_{\rm J}^{\rm th}$ (computed using numerically determined $\Delta\Sigma$ at the shock) varies with resolution, although relatively weakly, giving better agreement with simulations at higher resolution. Both the advective term and the time-dependent contribution shown in Figure \ref{fig:resolution}b are affected by varying resolution only weakly.

What does change significantly with resolution is the residual term ${\cal R}$ shown in Figure \ref{fig:resolution}c (it is multiplied by a factor of 10 for better representation). One can see that increasing resolution consistently reduces the magnitude of ${\cal R}$. This naturally brings us to the conclusion that the residual term is caused by the {\it numerical dissipation} due to the finite resolution of our simulations. 

The effect of the numerical residual is twofold. First, it gives rise to an unphysical contribution to the angular momentum balance in the disk, which corrupts the agreement between $\partial_R\left(R^2 T_{R\phi}\right)$ and $\partial_R F_{\rm J}^{\rm th}$. This effect is discussed further in \S \ref{sect:disc}. Second, higher numerical dissipation at lower resolution causes unphysical reduction of the density jump across the shock. Although small, this effect artificially decreases the value of $\partial_R F_{\rm J}^{\rm th}$ at lower resolution, which is seen in Figure \ref{fig:resolution}a.

\begin{figure*}
\centering
\includegraphics[width=1\textwidth]{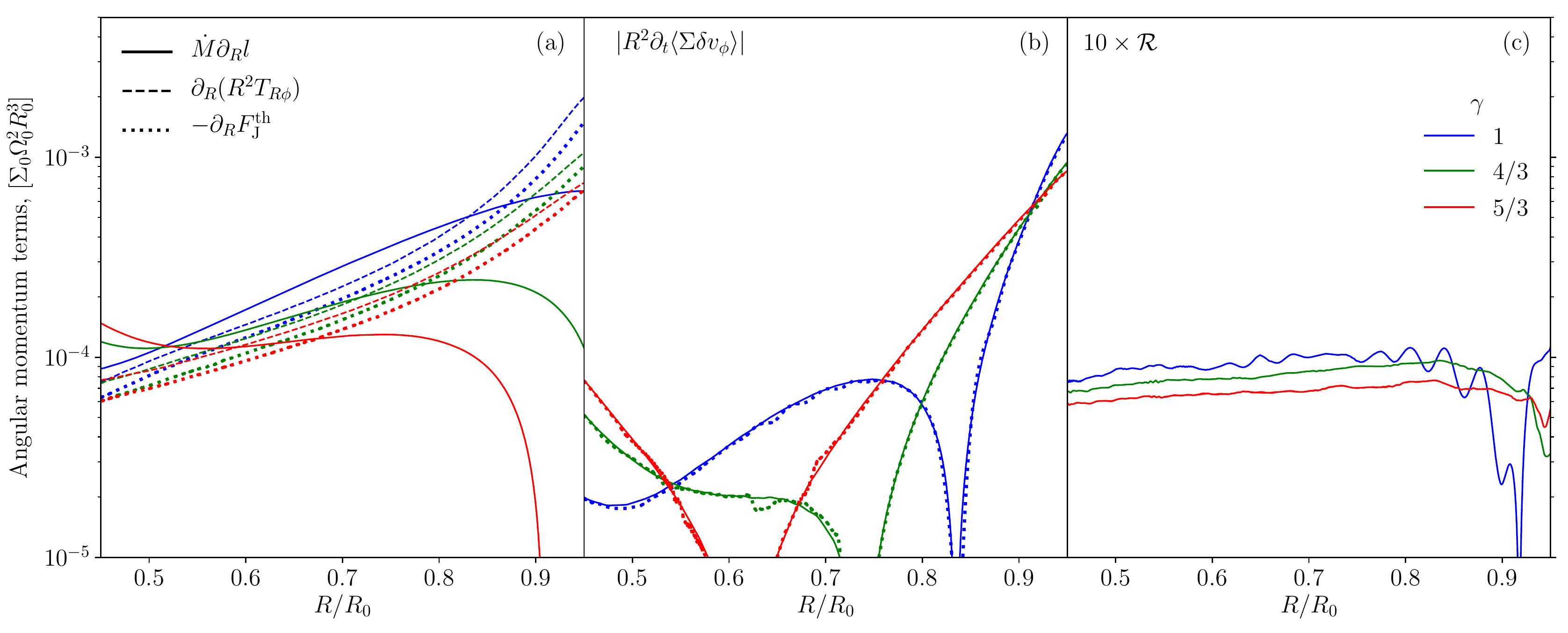}
\caption{
Same as Figure \ref{fig:AMterms}, but now showing the effect of varying the equation of state (we change the adiabatic index $\gamma$ as shown in the right panel) on the agreement between the theory and simulations. 
\label{fig:eos}}
\end{figure*}

Figure \ref{fig:residual} provides additional information on the behavior of the residual ${\cal R}$. It clearly shows that, everything else being equal (e.g. resolution, $H/R$, etc.), the value of ${\cal R}$ scales roughly {\it linearly} with the initial wave amplitude. This behavior is what one would expect from the usual viscous term in the fluid equations, reinforcing our interpretation of ${\cal R}$ as being due to numerical dissipation. Note that all physical shock-induced angular momentum contributions exhibit a significantly steeper dependence on $A_0$, see equations (\ref{eq:psiQ})-(\ref{eq:isoth}) and Figure \ref{fig:AMterms}a.

Figure \ref{fig:residual} also clearly shows how the residual decreases as the resolution goes up. At a fixed resolution ${\cal R}$ is also lower for smaller $H/R$, simply because the angular momentum flux carried by the wave of fixed amplitude is lower in colder disks, see Figure \ref{fig:hor}.


\subsection{Variation of the EOS.}  
\label{sect:EOS}


Equation of state (EOS) of the disk fluid should have an important effect for shock-mediated transport. This can be seen simply in the explicit dependence of the theoretical prediction (\ref{eq:psiQ}) on the adiabatic index $\gamma$ in the non-isothermal case. 

In Figure \ref{fig:eos} we compare the outcomes of three simulations with the same starting amplitude $A_0=1$ but having EOS with $\gamma=5/3$, $4/3$, and $1$ (isothermal). One can see that wave dissipation proceeds differently depending on the value of $\gamma$. Despite that, theoretical $\partial_R F_{\rm J}^{\rm th}$ correctly reproduces the variation of $\partial_R\left(R^2 T_{R\phi}\right)$ as $\gamma$ changes. Note that in non-isothermal runs we must use equation (\ref{eq:psiQ}), which apparently performs as well as the equation (\ref{eq:isoth}) does in the isothermal case.

\begin{figure*}
\centering
\includegraphics[width=1\textwidth]{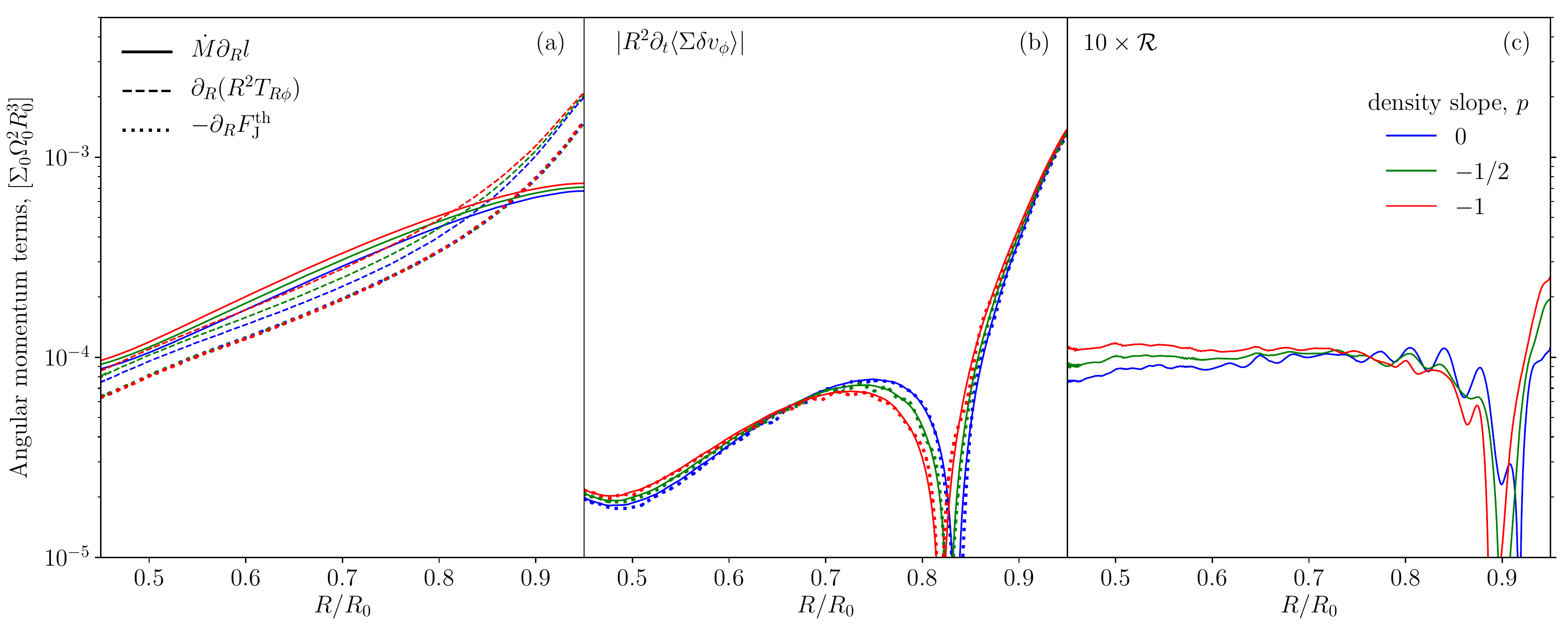}
\caption{
Same as Figure \ref{fig:AMterms} but now illustrating the effect of changing the slope of the background  surface density $p$, see equation (\ref{eq:Sprofile}), on the agreement between the theory and numerics. 
\label{fig:sig_slope}}
\end{figure*}

On the other hand, in our non-isothermal runs the advective term deviates from the stress term much stronger than in the isothermal case. This is a result of a more pronounced role of the time-dependent contribution (see Figure \ref{fig:eos}b), which is likely caused by the continuous and spatially non-uniform injection of entropy into the disk by the shock in the non-isothermal simulations. This steady growth of entropy (absent in the isothermal case) results in a more pronounced role of the time-dependent term in these runs.


\subsection{Variation of the background disk properties.}  
\label{sect:Sigma}


Most of the simulations shown in this paper feature radially uniform surface density profile (i.e. $p=0$). To demonstrate that this assumption does not affect our conclusions, in Figure \ref{fig:sig_slope} we show a series of runs with isothermal equation of state and outer amplitude $A_0=1$, in which we vary the power-law slope of the background density profile $p$ (which requires accounting for the radial pressure support in equation (\ref{eq:Omega})). 

\begin{figure*}
\centering
\includegraphics[width=1\textwidth]{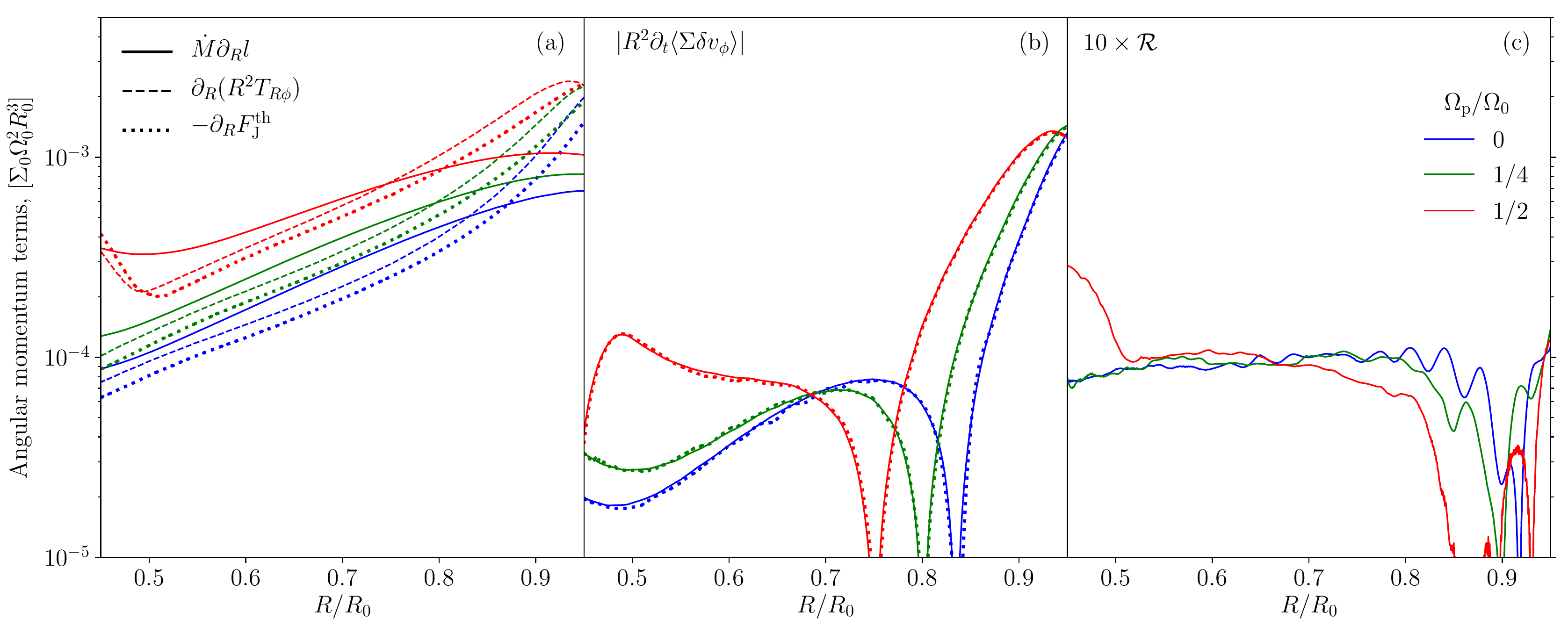}
\caption{
Same as Figure \ref{fig:AMterms} but now illustrating the effect of nonzero angular frequency $\Omega_{\rm p}$ of the imposed wave pattern on the agreement between the theory and numerics. Color scheme is illustrated in panel (c).
\label{fig:pat_rotation}}
\end{figure*}

One can see that the variation of $\Sigma(R)$ profile does surprisingly little to the numerical results, as both $\partial_R\left(R^2 T_{R\phi}\right)$ and the advective term exhibit only slight changes. The theoretical prediction varies even less. The agreement between $\partial_R\left(R^2 T_{R\phi}\right)$ and $\partial_R F_{\rm J}^{\rm th}$ remains good.

We also run several non-isothermal simulations in which we vary the radial profile of $c_{\rm s}$, while keeping $\Sigma(R)$ constant. Despite the  nonuniform temperature profile, we again found good agreement between the theoretical prediction (\ref{eq:dFJdr}) and the numerically determined $\partial_R F_{\rm J}$ (we do not provide the figure illustrating this). 


\subsection{Rotation of the perturbation pattern.}  
\label{sect:rot}


The final parameter that we varied in our simulations is the angular frequency $\Omega_{\rm p}$ of the imposed wave pattern. All of the simulations so far used non-rotating pattern, $\Omega_{\rm p}=0$, stationary in the inertial frame. In Figure \ref{fig:pat_rotation} we show the effect of a non-zero $\Omega_{\rm p}$ on our results, expressing $\Omega_{\rm p}$ in units of $\Omega_0$ --- Keplerian angular frequency at the outer edge of the disk. For these runs we impose perturbation at the outer boundary (the same as in the non-rotating case $\Omega_{\rm p}=0$), which uniformly rotates with $\Omega_{\rm p}/\Omega_0=0.25$ and $0.5$, allowing us to keep corotation outside the simulation domain.

One can see that variation of $\Omega_{\rm p}$ has a significant effect on the simulation results. The run with $\Omega_{\rm p}=0.5\Omega_0$ even exhibits the secondary shock (similar to that in Figure \ref{fig:hor} for $H_0/R_0=0.1$) starting at around $R\approx 0.5R_0$, resulting in the increase of $\partial_R\left(R^2 T_{R\phi}\right)$ inward of this radius. Nevertheless, despite these modifications, our analytical predictions (\ref{eq:dFJdr})-(\ref{eq:isoth}) still work well, and the agreement between the stress term and $\partial_R F_{\rm J}^{\rm th}$ for different values of $\Omega_{\rm p}$ in Figure \ref{fig:pat_rotation} remains very good.

\begin{figure}
\centering
\includegraphics[width=0.5\textwidth]{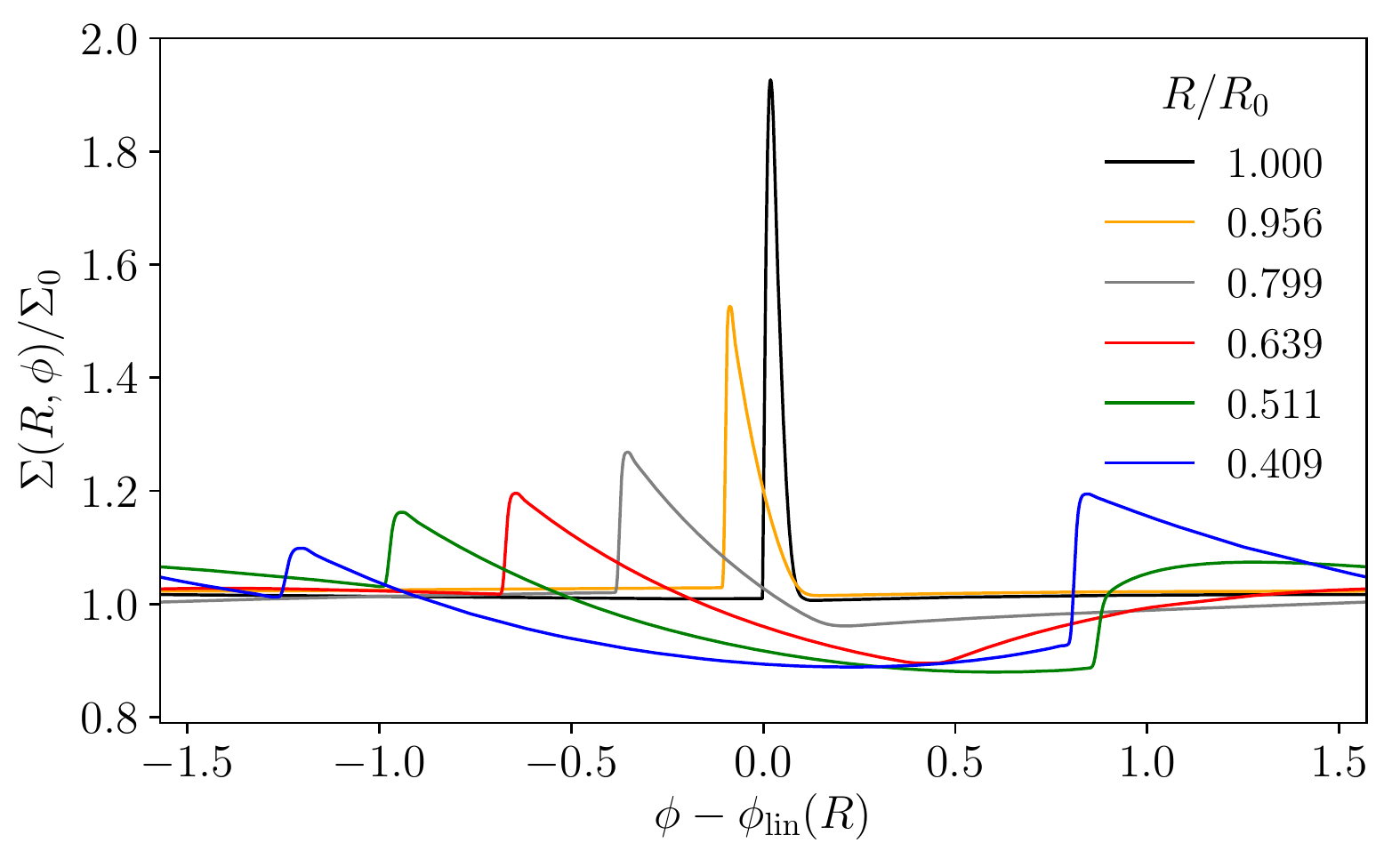}
\caption{Evolution of the wave profile, similar to Fig. \ref{fig:profile}, but now for a simulation with $H_0/R_0=0.1$. Faster nonlinear evolution of the wave in a colder disk results in the emergence of a secondary shock at $R\approx 0.5R_0$.  
\label{fig:shock_form}}
\end{figure}


\section{Secondary shocks.}
\label{sect:nontrivial}


One intriguing feature that emerges in some of our calculations is the {\it secondary shock}. It manifests itself in the angular momentum plots via the change in the behavior of $\partial_R\left(R^2 T_{R\phi}\right)$ with radius: the stress contribution, monotonically decaying with decreasing radius, suddenly starts to rise. This transition is clearly seen in our Figures \ref{fig:hor}a (for $H_0/R_0=0.1$) and \ref{fig:pat_rotation}a (for $\Omega_{\rm p}/\Omega_0=0.5$). Similar behavior has also been observed in simulations of \citet{Ryan}. 

In Figure \ref{fig:shock_form} we directly trace the appearance of this feature by showing the evolution of the wave profile in a simulation with $H_0/R_0=0.1$. One can see that as the primary wave propagates inwards, a region of the {\it negative} surface density perturbation gradually develops ahead of the main shock (which itself features $\Delta\Sigma>0$).  This behavior can be seen in the Figure \ref{fig:profile} as well. In both cases the right side of the negative $\Delta\Sigma$ segment of the profile gradually steepens due to the nonlinear effects \citep{GR01,Rafikov02}, until it breaks at $R\approx 0.5R_0$ in Figure \ref{fig:shock_form}. 

The reason why this does not happen in Figure \ref{fig:profile} can be related to a higher $H/R$. Indeed, it has been shown both locally \citep{GR01} and globally \citep{Rafikov02} that the rate at which nonlinear evolution of the weakly nonlinear density waves proceeds is set by the time-like coordinate 
\ba   
t\propto \left(\frac{H}{R}\right)^{-5/2}\Phi(R),
\label{eq:t}
\ea 
where $H/R$ is the aspect ratio at some fixed radius in the disk and $\Phi(R)$ is a function, monotonically increasing with the distance travelled by the wave, that absorbs the radial dependence of $t$ (its explicit form can be deduced using equations (32), (34) of \citet{Rafikov02}). For a given starting amplitude and shape of the wave, a shock forms when a certain critical value of $t$ is reached. It is clear from equation (\ref{eq:t}) that this happens earlier (i.e. after traveling a shorter radial distance $|R-R_0|$ from the outer boundary) in colder disks with lower $H/R$.

A very similar behavior has been seen in a number of recent studies of the disk-planet interaction \citep{Zhu,Fung15,Dong2gap,ArzZhuSt2017}, which generically demonstrate the emergence of a secondary shock at radii about halfway inward from the planetary orbit. In \citet{Bae} the secondary (or even tertiary) shocks were suggested to be the cause of the multiple ring-like structures in the protoplanetary disks, similar to those observed in HL Tau \citep{ALMA} and TW Hya \citep{Andrews}. The evolutionary sequence leading to secondary shocks in these studies is identical to the picture outlined above: the development of the negative $\Delta\Sigma$ segment of the wave profile, which gradually evolves into a shock due to nonlinear effects. The second stage is a robust phenomenon, analogous to the nonlinear development of the N-wave in \citet{GR01}. 

What is less clear is the origin of the negative $\Delta\Sigma$ segment of the wave profile (giving rise to this whole evolution) in the first place. \citet{Fung15} and \citet{Lee} suggested that the secondary shocks are excited by the planetary gravity at the ultraharmonic resonances. Our results cast doubt on this interpretation, or at least suggest that possible pathways to secondary shocks are not unique. 

Indeed, in our case we see the secondary shocks to appear in simulations where external gravitational potential is completely absent as the wave is driven simply by the imposed pressure perturbation sharply localized at the outer boundary of the disk. Moreover, most of our simulations feature non-rotating wave patterns, so resonances with the pattern frequency are also excluded as a culprit behind the secondary shocks. Finally, the radial location at which the secondary shock emerges in our simulations is certainly non-unique and depends on $H/R$, pattern speed of the perturbation $\Omega_{\rm p}$ and the initial wave amplitude. This is different from some previous studies, which typically find the secondary shock to appear at $\approx 0.6$ of the planetary semi-major axis \citep{Zhu}, although some dependence of the exact location on the disk temperature and planet mass was reported in \citet{Bae}. 

Unfortunately, at the moment we do not have a good theory explaining the emergence of the negative $\Delta\Sigma$ perturbations in our simulations. We observe this phenomenon in simulations with different initial amplitudes, including the lowest we tried ($A_0=0.0625$), which are essentially in the linear regime. This likely excludes the nonlinear effects, such as ultraharmonics \citep{Fung15,Lee}, from being responsible for the appearance of the negative $\Delta\Sigma$ feature in our simulations. Our best guess at the moment is that it may be somehow related to the way, in which we are imposing our boundary condition for the perturbation at the outer edge of the simulation domain. Future work will show whether this explanation holds.


\section{Time-dependent contribution.}  
\label{sect:time-dep}


Our simulations clearly demonstrate the important role played by the time-dependent term $R^2\partial_t\langle\Sigma \delta v_\phi\rangle$ in relating the angular momentum and mass transport, see equation (\ref{eq:AMconsShock}). In our case this term is particularly pronounced at large radii, close to the outer boundary of the simulation domain, where its contribution to the angular momentum balance is comparable to other terms, see Figures \ref{fig:AM_6.25}, \ref{fig:AMterms}b and others.

To better understand the nature of this term we note that the radial component of the Navier-Stokes equation can be written as $v_\phi^2R^{-1}\approx\Omega_{\rm K}^2 R+ \Sigma^{-1}\partial_R P$, where we neglected the inertial terms altogether. Given that our convention assumes that $v_\phi=\Omega_{\rm K} R+\delta v_\phi$, one then finds that (up to terms higher order in $\delta v_\phi$)
\ba
\delta v_\phi\approx\frac{1}{2\Omega_{\rm K}\Sigma}\frac{\partial P}{\partial R}
\label{eq:delta_v_phi}
\ea    
due to the radial pressure support in the disk. 

Plugging this expression into the definition of the time-dependent contribution to the angular momentum balance and azimuthally averaging, one finds
\ba
R^2\partial_t\langle\Sigma \delta v_\phi\rangle\approx \pi\frac{R^2}{\Omega_{\rm K}}\frac{\partial^2 \overline{P}}{\partial t\partial R},
\label{eq:time_dep}
\ea    
where $\overline{P}\equiv \langle P\rangle/2\pi$ is the azimuthally averaged pressure.

We show how well this approximation works in Figures \ref{fig:AMterms}b, \ref{fig:hor}b, \ref{fig:resolution}b, \ref{fig:eos}b-\ref{fig:pat_rotation}b. There we display (with dots) the behavior of the time-dependent term given by the equation (\ref{eq:time_dep}) using the radial profile of pressure measured in the simulations. Note that even though the majority of our simulations feature uniform {\it initial} profile of $\Sigma$ and radially-constant pressure, the time-dependent term is generally non-zero. This happens because even the weak modification of the $\Sigma(R)$ by the shock-driven radial redistribution of mass turns out being sufficient for the right hand side of the equation (\ref{eq:time_dep}) to provide a significant contribution. 

Figure \ref{fig:AMterms}b and other similar plots make it clear that the analytical approximation for the time-dependent term $R^2\partial_t\langle\Sigma \delta v_\phi\rangle$ works extremely well (the agreement is somewhat worse at lower resolution). For all wave amplitudes and at all radii the theoretical prediction falls right on top of the numerically determined values of the time-dependent term (the only mild exception being the lowest amplitude case $A_0=0.0625$ for which the numerical dissipation might play some role). This is a very reassuring finding. It implies, for example, that the knowledge of the radial profile of pressure allows us to accurately predict the radial profile of $\dot M$ using equations (\ref{eq:MdotGen}), (\ref{eq:dFJdr}), and (\ref{eq:time_dep}) even in non-steady disks. Another important application of this finding is discussed next.


\section{Shock-driven evolution of accretion disks.}  
\label{sect:formalism}


We are now in position to formulate a fully time-dependent equation for surface density evolution in a shock-mediated disk. First, using our result (\ref{eq:MdotGen}), the continuity equation $\partial_t\overline{\Sigma}=(2\pi R)^{-1}\partial_R \dot M$ describing the evolution of the azimuthally averaged disk surface density $\overline{\Sigma}$ can be written quite generally as 
\ba     
\frac{\partial\overline{\Sigma}}{\partial t} &+& \frac{1}{2\pi R}\frac{\partial}{\partial R}\left[ \left(\partial_R l \right)^{-1} \left(\partial_R F_{\rm J}-R^2\partial_t\langle\Sigma \delta v_\phi\rangle\right)\right]
\nonumber\\
&=& 0.
\label{eq:cont}
\ea 

Second, equation (\ref{eq:time_dep}) allows us to put this expression in a closed form:
\ba    
\frac{\partial\overline{\Sigma}}{\partial t} &+& \frac{1}{2\pi R}\frac{\partial}{\partial R}\left[ \left(\partial_R l\right)^{-1}\left(\partial_R F_{\rm J}-\pi\frac{R^2}{\Omega_{\rm K}}\frac{\partial^2 \overline{P}}{\partial t\partial R}\right)\right]
\nonumber\\
&=& 0.
\label{eq:cont_final}
\ea    
Note that we can set $l=l_{\rm K}=\Omega_{\rm K} R^2$ in this equation without loss of accuracy. Indeed, one can show that the extra term inside the square brackets arising from the deviation of $l$ from $l_K$ due to the non-zero $\delta v_\phi$ can be comparable to the last term (due to the time-dependent pressure support) only when the disk evolves slowly, on viscous timescale. However, in that case both the correction to $\partial_R l$ and the last term are negligible compared to $\partial_R F_{\rm J}$ by a factor of at least $\sim (H/R)^2$ anyway.

Together with the analytical prescription (\ref{eq:dFJdr}) for $\partial_R F_{\rm J}$ the equation (\ref{eq:cont_final}) represents a closed form evolution equation for the (azimuthally averaged) surface density $\overline{\Sigma}$. Disk surface density enters this equation not only through the first, explicitly time dependent, term and $\partial_R F_{\rm J}$, but also through the dependence of $\overline{P}$ upon $\overline{\Sigma}$ in the last term. Equation (\ref{eq:cont_final}) thus represents a fully self-consistent framework for following the evolution of $\overline{\Sigma}$ in the shock-mediated accretion disks, provided that both the thermodynamic state of the disk (i.e. the radial behavior of the sound speed $c_{\rm s}$) and the radial profile of the shock strength (i.e. $\Pi(R)$) are fully specified.

Shock-mediated evolution of an accretion disk should eventually drive its inner regions towards a (quasi-)steady state (far from the corotation region corresponding to the pattern frequency $\Omega_{\rm p}$). In this case one can set $\partial_t=0$ in equation (\ref{eq:cont_final}), resulting in $\dot M={\rm const}$. Equations (\ref{eq:MdotGen}) and (\ref{eq:dFJdr}) then imply that the surface density profile in this part of the disk should converge to
\ba    
\overline{\Sigma}(R) =\frac{\dot M}{2m}\frac{\Omega(R)}{c_{\rm s}^2(R)\psi_Q[\Pi(R)]},
\label{eq:steady}
\ea    
where we assumed Keplerian rotation. This expression illustrates that in steady state the radial behavior of the disk surface density is fully determined by the radial profiles of the background temperature (i.e. $c_s$) and of the pressure jump at the shock $\Pi$.

Note that equation (\ref{eq:cont_final}) is not limited to shock-mediated disks, but applies to other types of disks as well. Recalling that in general $\partial_R F_{\rm J}=\partial_R\left(R^2 T_{R\phi}\right)$ one could apply this equation to disks evolving due to other mechanisms of the angular momentum transport, for which the stress tensor $T_{R\phi}$ can be specified (e.g. some kind of effective viscosity, MRI, and so on).

Equation (\ref{eq:cont_final}) is different from the classical form (without the time-dependent term) derived in \citet{lynden-bell_1974}, which is broadly used to study one-dimensional viscous disk evolution. This is because in their derivation \citet{lynden-bell_1974} neglected pressure support altogether and set the specific angular momentum of the disk fluid to be equal to the Keplerian value. This is fully legitimate if the disk evolves slowly, e.g. on global viscous timescale. However, our simulations demonstrate that in more rapidly evolving disks (e.g. due to initial readjustment of the disk structure to the shock-imposed torque, as in our case) the time-dependent term may not be neglected and should be properly accounted for. It is likely that this term plays important role in various astrophysical disks experiencing rapid evolution, e.g. during outburst events.


\section{Discussion}  
\label{sect:disc}


Results of simulations presented in \S \ref{sect:results} and \ref{sect:theory} clearly support theoretical calculation of the angular momentum deposition by the spiral shocks of arbitrary strength presented in \citet{Rafikov16}. In all cases when the shock amplitude is high and angular momentum injection by the wave (i.e. the stress term) dominates over the numerical effects (residual ${\cal R}$) we find good agreement between $\partial_R\left(R^2 T_{R\phi}\right)$ and $\partial_R F_{\rm J}^{\rm th}$. The agreement is worse in simulations with small wave amplitudes, but even then the analytical prescription of \citet{Rafikov16} definitely reproduces at least the qualitative behavior and the magnitude of $\partial_R F_{\rm J}$.

Our results strongly emphasize the importance of high resolution that the simulations trying to account for the disk shock phenomena must have. As demonstrated in Figure \ref{fig:AM_6.25}, at low wave amplitudes the physical contributions to the angular momentum balance become comparable with the numerical residual term (or even subdominant). At high wave amplitudes, in the fully nonlinear regime, situation improves because the magnitude of the physical angular momentum contributions grows faster than ${\cal R}$ (Fig. \ref{fig:residual}). This likely explains why \citet{Ju} saw the numerical residual to be close to zero in their study of high-amplitude spiral shocks in disks of cataclysmic variables. 

\begin{figure*}
\centering
\includegraphics[width=0.5\textwidth]{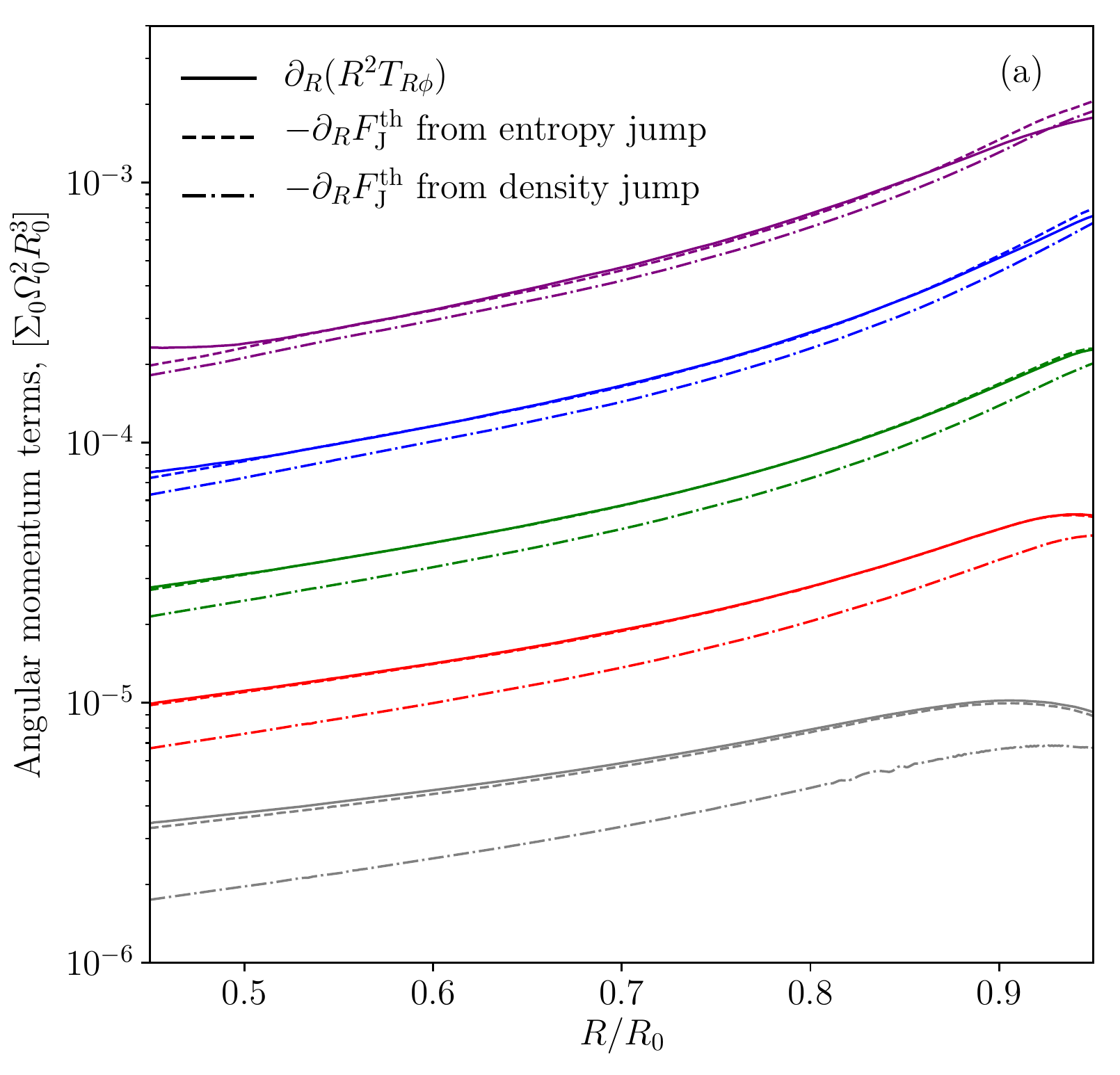}~\includegraphics[width=0.5\textwidth]{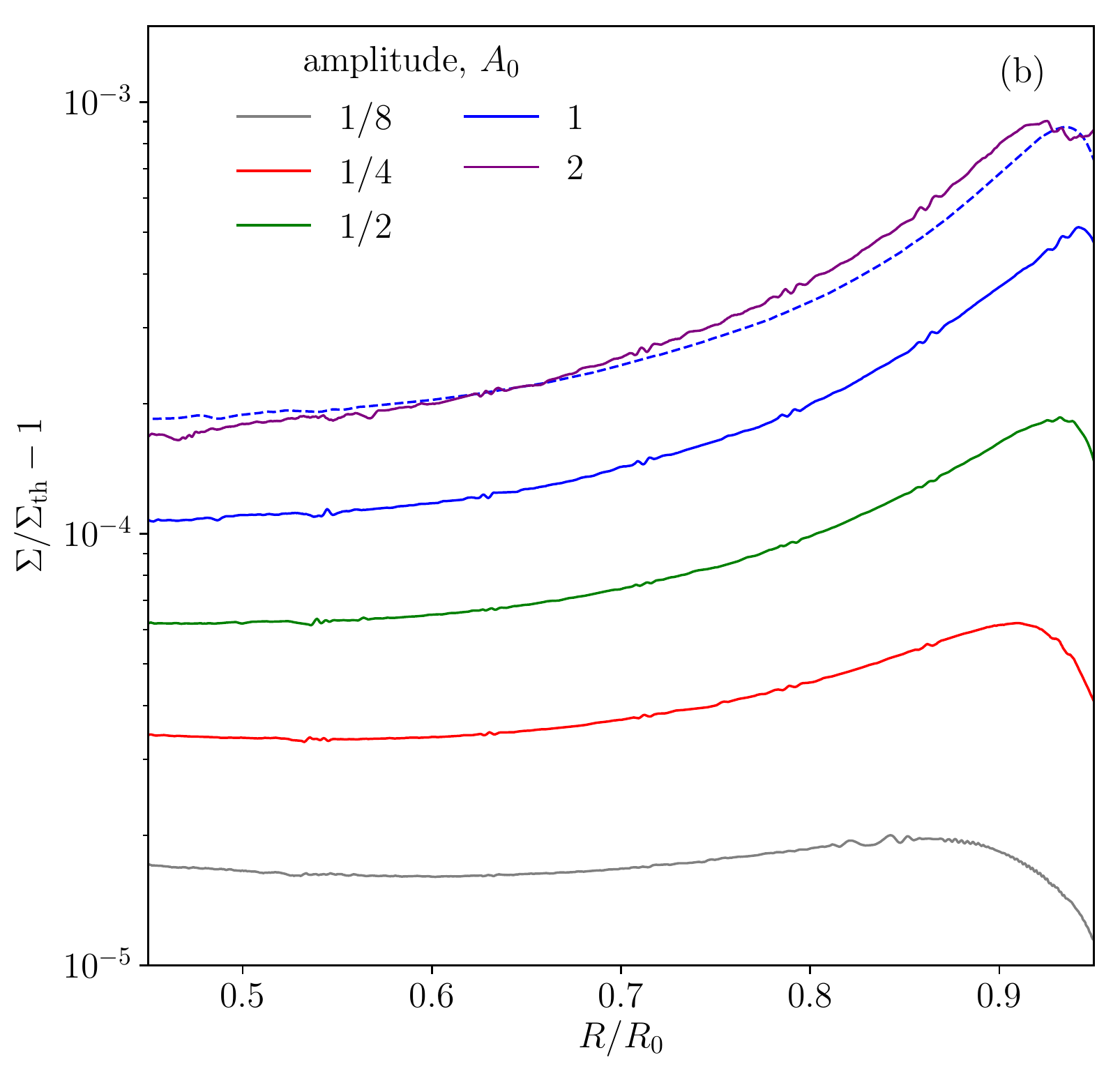}
\caption{
(a) Solid, dashed and dot-dashed lines represent the stress term $\partial_R (R^2 T_{R\phi})$, and the theoretical predictions for $\partial_R F_{\rm J}^{\rm th}$ computed using entropy jump (\ref{eq:psiQ_entr}) and density jump (\ref{eq:psiQ_rho}), correspondingly. (b) The difference between the density jump measured from simulation and theoretical prediction (\ref{eq:rho_jump}) computed from the measured pressure jump. The blue dashed line on this figure represents the run with two times lower resolution (as compared to the blue solid line). Both figures use the same color scheme.
\label{fig:entropy}}
\end{figure*}

It is also clear from Figure \ref{fig:AMterms} that in most cases the residual term $\mathcal{R}$ cannot fully explain the difference between the stress term $\partial_R\left(R^2 T_{R\phi}\right)$ and theoretical $\partial_R F_{\rm J}^{\rm th}$. We believe that this difference is also caused by numerical effects. In Figure \ref{fig:entropy}a we again show the angular momentum terms for simulations with adiabatic equation of state ($\gamma = 5/3$) and different wave amplitudes. Solid line shows the stress term, while other lines represent two kinds of the theoretical prediction for the stress term: one computed from the density jump (\ref{eq:psiQ_rho}) and shown with dot-dashed line, and another computed via the entropy jump at the shock front (dashed curve), as suggested in \citet{Rafikov16} and previously implemented in \citet{Ryan}. The latter can be written in terms of the density and pressure jumps at the shock as
\begin{equation}
\psi_Q = \frac{1}{\gamma-1} \left[\Pi (\Sigma/\Sigma_{\rm ps})^{-\gamma} - 1\right].
    \label{eq:psiQ_entr}
\end{equation}
We see surprising disagreement between the theoretical predictions computed using two methods. On the other hand, the angular momentum flux computed using entropy jump reproduce numerical stress term very well. 

The only difference between equations (\ref{eq:psiQ_rho}) and (\ref{eq:psiQ_entr}) is the connection between the pressure and density jumps \citep{LandauLifshitz1959}:
\begin{equation}
\frac{\Sigma_{\rm th}}{\Sigma_{\rm ps}} = \frac{\gamma - 1 + (\gamma+1)\Pi}{\gamma+1 +(\gamma -1 )\Pi}.
    \label{eq:rho_jump}
\end{equation}
Figure \ref{fig:entropy}b shows that our simulations do not reproduce this relation exactly, i.e. the actual post-shock density $\Sigma$ is slightly different from $\Sigma_{\rm th}$. Although the error is very small, of the order of $10^{-4}$, it causes a big effect on $\psi_Q$. One can estimate the error in $\psi_Q$ as
\begin{equation}
    \delta\psi_Q \approx \frac{\gamma}{\gamma-1} \Pi \left(\Sigma_{\rm th}/\Sigma_{\rm ps}\right)^{-\gamma} (\Sigma/\Sigma_{\rm th} - 1).
    \label{eq:diff_psiQ}
\end{equation}
We verified that the difference between the two methods of analytical evaluation of the shock-induced stress can indeed be approximated by equation (\ref{eq:diff_psiQ}) with very good accuracy. The theoretical angular momentum flux\footnote{We do not show it in Figure \ref{fig:entropy}a to avoid confusion.} computed using density jump and corrected using equation (\ref{eq:diff_psiQ}) coincides with $\partial_R F_{\rm J}^{\rm th}$ computed using entropy jump almost perfectly.  

Blue dashed line in Figure \ref{fig:entropy}b shows the deviation in the post-shock density at the same shock amplitude as the blue solid line ($A_0=1$), but measured in a simulation with two times lower resolution. One can see that lowering resolution by a factor of two almost doubles the error in density jump. This exercise demonstrates that the difference between the two methods of computing theoretical angular momentum contributions due to shocks is indeed numerical. 

One possible explanation for this effect might come from the fact that shocks in our simulation are not true discontinuities. Shock front is in fact smeared over several grid points, making pinpointing exact boundaries of the shock region difficult. If our shock detection algorithm (see \S \ref{sect:shock}) makes a slight spatial error in the shock position, the fluid might expand adiabatically over this small distance. This will not affect the change of entropy (explaining why the calculation based on entropy jump is robust), but will cause the relationship between the density and pressure jumps to deviate from the theoretical expectation (\ref{eq:rho_jump}). If that were the reason, then we would expect the error in jump condition to go down with resolution, and to go up with $A_0$ as the number of grid points needed to capture the shock increases with its amplitude. Both trends are indeed observed in our simulations.

It is also worth reiterating that predicting the shock-driven mass accretion rate $\dot M$ based on the shock strength is in general less trivial than predicting the angular momentum deposition rate. A direct connection between $\partial_R F_{\rm J}^{\rm th}$ and $\dot M$ established in \citet{Rafikov16} is valid only in steady state disks. Their relationship becomes more complicated in disks experiencing rapid evolution, as we demonstrated in \S \ref{sect:results} and \ref{sect:theory}. Nevertheless, by carefully examining individual contributions to the angular momentum balance we can understand how the time variability of the disk properties can be accounted for (see \S \ref{sect:time-dep}) and provide an explicit prescription for computing $\dot M$, see equations (\ref{eq:MdotGen}) and (\ref{eq:time_dep}).

The general one-dimensional framework for the disk evolution represented by equation (\ref{eq:cont_final}) should be useful for other studies of shock-mediated disks (as well as for accretion disks evolving due to internal stresses). It allows for fast and accurate exploration of many evolutionary models of such disks, avoiding computationally expensive multi-dimensional simulations (which could also be contaminated by the numerical artifacts).

The disk setup used in this work (both numerical and analytical) is explicitly  two-dimensional. However, we do not expect our results to change in the case of a fully three-dimensional disk, as long as one is able to adequately describe the radial and vertical behaviors for the shock strength. In that case, like in \citet{Rafikov16}, one could still use  energy and angular momentum conservation to provide a connection between the entropy production at the shock and the angular momentum deposition rate. This connection is also independent of the cooling mechanism of the disk, as passage of gas through the shock is essentially instantaneous. Cooling behavior should affect only the background temperature of the disk, which can be accounted for by explicitly considering disk thermodynamics. 

Our study also neglects the effect of magnetic fields, which modify the jump conditions at the shock and should affect the calculation of $\psi_Q$. However, as long as magnetic fields remain sub-thermal (which is almost always the case for the system we focus on), their effect on the shock jump conditions and entropy calculation should be subdominant. Angular momentum transport driven by magnetized turbulence through MRI (if it operates in the disk) can be easily accounted for in the framework of our approach by simply adding corresponding stress term in equations (\ref{eq:cont}) and  (\ref{eq:cont_final}).

Note that throughout our study we did not specify the driving agent for the spiral wave. In many astrophysical disks it could be caused by a massive perturber --- a planet or a binary companion \citep{Dong_spiral,Dong_star,Zhu,ArzZhuSt2017}. In that case our results would apply to the {\it coasting} phase of the wave propagation, far from the perturber, when it is no longer significantly affected by its gravity. 

Alternatively, recently \citet{Montesinos} proposed that spiral waves may be driven by the non-uniform illumination of the protoplanetary disks by the central star, caused e.g. by the misaligned inner disk. This possibility was invoked to explain the non-axisymmetric features seen in the disks of HD 142527 \citep{Casassus,Marino} and HD 100546 \citep{Benisty16}. Our results would apply equally well to spiral waves sourced by this or any other mechanism (e.g. gravitational instability, see \citet{Meru}). 


\subsection{Comparison with the previous work}  
\label{sect:compare}


Several numerical studies have recently explored the effect of global spiral shocks on disk evolution. \citet{Ryan} directly verified the analytical predictions of \citet{Rafikov16} for the spiral shock-driven angular momentum transport in mini-disks around the components of the supermassive black hole binaries. There are several differences between their work and ours. First, \citet{Ryan} run their simulation in full general relativity, while we use Newtonian gravity. Second, to characterize the angular momentum injection by the wave they used the entropy jump at the shock, as suggested in \citet{Rafikov16}. In our present work we use the density jump at the shock front, which is easier to determine observationally, for that purpose. Third, \citet{Ryan} run their simulations until the disk reached a steady state, so that the time-dependent contributions were small in their case. Our simulations are not run for nearly as long, forcing us to explicitly account for the time-dependent terms, see \S \ref{sect:time-dep}.

\citet{Ju,Ju2017} studied the role of spiral shocks in driving accretion in disks of cataclysmic variables, both in hydro and MHD setups. Our analysis of the different torque contributions is very similar to that in \citet{Ju}. Despite the improved numerical resolution, we see our results significantly affected by numerical dissipation, especially at low amplitudes (see Figure \ref{fig:AM_6.25}b), compared to the calculations in \citet{Ju}. Presumably, the lack of noticeable residual ${\cal R}$ is their simulations is caused by the higher amplitude of the spiral shocks driven by a massive companion.

\citet{Ju2017} also found that the shock-driven transport becomes inefficient when the disk aspect ratio $H/R$ is reduced. This is consistent with our results in \S \ref{sect:h-r}, where we show that the angular momentum deposition by the spiral shock decays with the distance traveled by the wave considerably faster in disks with lower $H/R$, see Figure \ref{fig:hor}. The accelerated decay is caused by the more rapid nonlinear evolution and associated damping of the waves propagating in colder disks, see the discussion around equation (\ref{eq:t}). As a result, we  naturally expect the spiral waves reaching central regions of the disks in cataclysmic variables (which have $H/R\sim 10^{-2}$) to have very low values of the pressure jump $\Pi$ at the shock, rendering them inefficient at transporting angular momentum far from the disk edge. Protoplanetary \citep{Rafikov16} and circumplanetary \citep{ZhuCP} disks, which have higher $H/R$, may provide better environments for the shock-driven transport.


\section{Summary}  
\label{sect:summary}


We have carried out a numerical study aimed at understanding the effects of the global spiral shocks on astrophysical disks. The main goal of this work was verification of the analytical predictions of \citet{Rafikov16} for the angular momentum deposition in the disk by the global shocks. Our main results can be summarized as follows.

\begin{itemize}

\item Our numerical results for the angular momentum deposition rate due to disk shocks show good agreement (at the level of tens of per cent) with the predictions of \citet{Rafikov16}, evaluated using numerically determined shock strength. This agreement persists as we vary different relevant physical parameters --- initial amplitude of the density wave, aspect ratio of the disk, equation of state of the disk fluid, radial profiles of the background surface density and temperature, angular frequency of the perturbation pattern (\S \ref{sect:theory}).

\item High resolution is essential for fully capturing the effect of global shocks on the disk structure, and we quantify the effect of numerical dissipation on our simulations (\S \ref{sect:res}). Properly accounting for the detrimental effects of numerical dissipation is especially important for low amplitude shocks.

\item We generalize the relationship between the shock-driven angular momentum dissipation and the mass accretion rate $\dot M$, derived in \citet{Rafikov16} for steady disks, to cover the situations when the disk evolves rapidly. In this case the time-dependent terms provide important contribution to the angular momentum budget and must be explicitly accounted for.

\item We provide a simple analytical prescription for calculating the behavior of these time-dependent terms using the knowledge of the radial pressure profile in the disk (\S \ref{sect:time-dep}). This prescription is found to be in excellent agreement with the numerical results.

\item These findings allow us to formulate a closed form equation for the shock-driven evolution of accretion disks, valid even when the disk properties change on time scales shorter than the ``viscous" timescale (\S \ref{sect:formalism}). This framework can be used for accurate and efficient modeling of the shock-driven disk evolution in one (radial) dimension.

\end{itemize}

Successful confirmation of the analytical prescription for the shock-driven disk evolution obtained in \citet{Rafikov16} opens a way for semi-analytical studies of the long-term disk evolution driven by the spiral shocks, once the radial profile of the shock strength (i.e. density or pressure jump at the shock) is specified. Fully self-consistent and complete description of the evolution of the density wave as it propagates through the disk (and damps), which is needed for prescribing the shock amplitude at every radius, is a subject of future work.

\acknowledgements

We thank James Stone for helpful discussions and comments, as well as for making Athena++ publicly available. Financial support for this study has been provided by NSF via grants AST-1409524,  AST-1515763, and NASA via grants 14-ATP14-0059, 15-XRP15-2-0139. All simulations were carried out using facilities supported by the Princeton Institute of Computational Science and Engineering.



\bibliographystyle{apj}
\bibliography{references}

\end{document}